# Consequential LCA for territorial and multimodal transportation policies: method and application to the free-floating e-scooter disruption in Paris

Anne de Bortoli[1*], Zoi Christoforou[1]

**1** Department of Civil Engineering, University of Patras, Panepistimioupoli Patron 265 04, Patras, Greece

\* Corresponding author; e-mail: anne.de-bortoli@enpc.fr

ABSTRACT

The indirect environmental impacts of transportation disruptions in urban mobility are frequently overlooked due to a lack of appropriate assessment methods. Consequential Life Cycle Assessment (CLCA) is a method to capture the environmental consequences of the entire cause and effect chain of these disruptions but has never been adapted to transportation disruption at the city scale. This paper proposes a mathematical formalization of CLCA applied to a territorial mobility change. The method is applied to quantify the impact on climate change of the breakthrough of free-floating e-scooters (FFES) in Paris. An FFES user survey is conducted to estimate the modal shifts due to FFES. Trip substitutions from all the Parisian modes concerned are considered – personal or shared bicycles and motor scooters, private car, taxi and ride-hailing, bus, streetcar, metro and RER (the Paris metropolitan area mass rapid transit system). All these Parisian modes are assessed for the first time using LCA. Final results estimate that over one year, the FFES generated an extra thirteen thousand tons of $CO_2$eq under an assumption of one million users, mainly due to major shifts coming from lower-emitting modes (60% from the metro and the RER, 22% from active modes). Recommendations are given to enhance their carbon footprint. A scenario analysis shows that increasing the lifetime mileage is insufficient to get a positive balance: reducing drastically servicing emissions is also required. A sensitivity analysis switching the French electricity mix for eleven other country mixes suggests a better climate change effect of the FFES in similar metropolitan areas with higher electricity carbon intensity, such as in Germany and China. Finally, the novelty and the limits of the method are discussed, as well as the results and the role of e-scooters, micromobility, and shared vehicles towards sustainable mobility.

# 1 Introduction

In summer 2018, Free-Floating E-scooters (FFES) started to be deployed in Paris by the leading company of the market (Kristanadjaja, 2019). After 6 months, the company reported two million rides made by 315 000 users in the city, representing almost 8% of its annual rides worldwide (Lime, 2019). One year after this launch, the city counted thirteen operators and a total fleet estimated between 20 000 and 40 000 FFES (Cosnard, 2019), making Paris the first market for FFES in the world (6t, 2019). The rapid and massive breakthrough of these microvehicles in a fragile urban traffic status quo has triggered resentments from other street users, and sometimes a mistrust from public authorities. In France, the Government developed a new section in the Highway Code to regulate motorized microvehicles, now strictly forbidden on sidewalks or at speeds over 25 km/h – 15 mph (French Department of Homeland Security, 2019). The city of Paris decided to launch an invitation to tender based on environmental and social criteria to select two or three companies allowed to operate (Cosnard, 2019). But the environmental consequences of FFES are highly uncertain and have just started being assessed.

The FFES has often been advertised in their operator communication campaigns as an opportunity to preserve the environment, as it would reduce congestion, air pollution, automobile usage, and, most of all, greenhouse gas (GHG) emissions (Lime, 2018). Some figures about the effect on climate change have been given, without information on the assessment methodology. The leading company has announced in its one-year activity report that its first 6 million rides would have saved 2.3 M $kgCO_2eq$, with around 14% of these savings occurring in Seattle, Washington, and 10% in Dallas, Texas (Lime, 2018). Nevertheless, no similar figures were presented in the following year's annual report (Lime, 2019).

Indeed, the potential environmental benefit of FFES was severely contested in 2019 through the question of the FFES lifespan: a first study quickly estimated it equal to 29 days in Louisville, Kentucky (Quartz, 2019), based on an ambiguous FFES unique identification number from an open dataset formerly provided by the city (Mattingly, 2019). Without information on the distance traveled a day, this study does not provide any clue to the environmental performance of FFES. Nevertheless, a first life cycle assessment (LCA) attempt on FFES was performed by a consultant, using generic data and rough assumptions (Chester, 2019). Chester found that FFES would emit between 200 and 460 $gCO_2eq/km$. More recently, four personal microvehicles have been assessed using LCA in the context of Paris. This study estimated the impact of the personal e-scooter (ES) as low as 12 $gCO_2eq/km$, provided a long lifetime mileage of 15 600 km, based on the maximum lifespan reported by users (de Bortoli et al., 2020c). Finally, Hollingsworth et al. conducted an LCA of FFES in the context of Raleigh, North Carolina, including uncertainty analysis. Based on limited local surveys and a material inventory obtained from the dismantling of a Xiaomi M365, they found very variable FFES carbon footprints depending on the assumptions made, and an average impact in Raleigh's context around 126 $gCO_2eq/km$ (Hollingsworth et al., 2019). According to this study, half of the impact would come from the manufacture of the microvehicle, mainly from aluminum and the lithium-ion battery. In this study though, the aluminum alloy carbon footprint was reported twice as important as in the selected database - 13 $kgCO_2eq/kg$ based on the article supplementary material figures, instead of around 6.5 $kgCO_2eq/km$ in ecoinvent V3.3 – potentially suggesting an over-estimation of the manufacturing stage impact. The rest of the FFES impact would mainly come from servicing, i.e. the collection and distribution of FFES to charge the batteries.

These studies give a broad environmental performance range of ES and first LCA models that must be adapted to each specific situation - to match the technosphere and the ES usage characteristics - to get a robust environmental comparison of ES and other means of transportation. But to our knowledge, no assessment of the global environmental impact of the FFES breakthrough at a regional scale has been performed yet. In particular, the FFES are mostly used instead of other pre-existing transportation modes, and extra trips are also generated by this new transportation service, as demonstrated by different surveys (6t, 2019; Hollingsworth et al., 2019; Lime, 2018; Portland Bureau of Transportation, 2018). Finally, the total environmental effect of this change in mobility consumption has not been assessed and is not properly understood.

To design transportation policies, the carbon footprint of mobility changes is often assessed on a truncated perimeter, only considering tail-pipes emissions. This approach can lead to misleading findings. The most complete studies use attributional LCA. LCA is a method to calculate the environmental performance of a system over its entire life cycle, from cradle to grave. It is standardized by the ISO 14040 and 14044 (AFNOR, 2006a, 2006b), these standards still allowing great flexibility in the way of using the method. Different approaches in LCA can be adopted. Attributional LCA (ALCA) is often opposed to consequential LCA (CLCA). ALCA allows for calculating the environmental impact of a system, e.g. a product or a service. It is most common and was widely applied to transportation components – to infrastructures and vehicles first, then to means of transportation (de Bortoli et al., 2020b). Later, it was used at the regional scale to assess the environmental impacts of a static mobility demand (Le Féon, 2014), and a dynamic mobility demand depending on mobility supply when coupled with an activity-based model (Baustert et al., 2019). But while ALCA rather intends to assess the average environmental impact of a static system, CLCA must be used to assess the environmental consequences of an action or a decision on a system. CLCA particularly includes economic market mechanisms into the analysis (Zamagni et al., 2012). In particular, it considers how a change in demand affects activities (Weidema et al., 2013) and thus environmental impacts, e.g. how a rising demand in electricity can change the technology mix. This is why CLCA should be systematically used when a social responsibility paradigm is addressed by the assessment (Weidema et al., 2018), e.g. for transportation public policies. FFES have triggered an intense social debate about their value for society, and micromobility embodies a change in the mobility paradigm. Especially, switching from conventional modes to FFES rides implies a change in demand, thus many industrial production changes, e.g. in terms of energy or material consumptions. Thus, a consequential approach must be selected to calculate the environmental impact of the FFES disruption.

Nevertheless, CLCA remains rare in the transportation sector. The CLCA concept appeared in 1993 (Weidema, 1993). It has been discussed and framed out since then, but no consensus has been achieved, and it is still rarely and heterogeneously applied (Zamagni et al., 2012). Sanden and Karlstrom first used it on bus fuel cell technology (2007). Then, Spielmann et al. applied it to high-speed transportation technologies and demonstrated the crucial effects of demand changes on the environmental performance of transportation projects (2008). Numerous applications for biofuels have also been performed - starting by the seminal study from Reinhard and Zah (2009) - as biofuel production implies heavy cross-sectoral systemic effects than must be addressed consequentially. More recently, a project of bus rapid transit in a Parisian suburb has been assessed hybridizing CLCA and the French socio-economic appraisal, based on an attributional background data set but considering modal shifts (de Bortoli, 2016). However, to our knowledge, no CLCA method applied to regional mobility and technological disruption has been properly conducted and mathematically formalized so far.

To fill this gap, we propose in this paper a framework for complete consequential LCA applied to integrated transportation changes, with a full mathematical formalization, including how to switch from classical survey trip-based data to kilometer-based data. Thanks to this method, we can properly assess and discuss the questionable environmental impact of ES. We will especially use it to calculate the early evolution in the GHG emitted at the city scale due to the FFES disruption over one year in Paris, as a hotspot for FFES. First, the generic method and its extended application to the case study will be explained, before detailing results, performing sensitivity analyses to get a broader sense of FFES environmental performances, and discussing the method and the case study results.

## 2  Method

The method is based on modal shift data and CLCA.

### 2.1  Consequential LCA for mobility: generic equations

**Modal shift calculation**

The first objective is to access the vector $\Delta PKT_{ms}=(\delta pkt_i)$ of the modal shift mileage disaggregated by mode on the analysis period t. We will use the general equation 1 to calculate its terms $\delta pkt_i$, with $i$ the transportation mode considered, $PKT_i$ the number of passenger-kilometers traveled (pkt) over the period t by the mode $i$ in the reference scenario (no transportation disruption), $PKT_i$' the number of pkt under the alternative scenario, and $\Delta PKT_i$ the result.

$$\delta pkt_i = PKT_i' - PKT_i = \Delta PKT_i \qquad (1)$$

**Environmental impact calculations**

Then, the terms $ei_i$ of $EI$ - the vector of the environmental impact from the modal shift that has occurred over the period $t$ - must be calculated following the equation 2, with $EF_i$ (resp. $EF_i'$) the environmental factor of the mode $i$ under the reference scenario (resp. the alternative scenario). The total environmental impact is then calculated summing the terms $ei_i$ of $EI$.

$$ei_i = EF_i'.PKT_i' - EF_i.PKT_i = EF_i.\Delta PKT_i \qquad (2)$$

Calculating equation 2 requires estimating the environmental factor $ei_i$ of each mode $i$ affected by the arrival of FFES in Paris, an environmental factor being a unitary environmental impact, here the unit being per pkt. If the environmental factor does not vary significantly over time, the calculation can be simplified as followed by the last term of the equation, using an approximate $EF_i$ factor. To calculate the equation 2, we use equation 3 that presents the formula to calculate the environmental impact of a mode $i$ with a vehicle-infrastructure integrated approach, with $EF_{veh,i}$ the environmental factor from the vehicle component of the mode $i$ per pkt, $EF_{infra,i}$ the environmental factor from the infrastructure component of the mode $i$ per pkt, $EF_{1veh,i}$ the environmental factor of one average vehicle of the mode $i$ over its life cycle, $PKT_{1veh,i}$ the lifetime mileage of one vehicle, $PKT_{ij}$ the number of passenger-kilometers traveled on the infrastructure $j$ for the mode $i$, $a_{i,j}$ the "specific infrastructural demand" (Spielmann et al., 2007) of type $j$ by the mode $i$ (see further explanations below), $q_j$ the number of units of the infrastructure $j$ used by the mode $i$, $EF_{1u,j}$ the environmental factor of one unit (surface or length) of the infrastructure $j$, $b_{ij}$ the infrastructural allocation factor of the infrastructure $j$ for the mode $i$, and $VKT_{ij}$ the number of vehicle-kilometers traveled (vkt) by mode $i$ on the infrastructure $j$.

$$EF_i = EF_{veh,i} + EF_{infra,i} = \frac{EF_{1veh,i}}{PKT_{1veh,i}} + \sum_j \frac{1}{PKT_{ij}}.a_{ij}.q_j.EF_{1uj} = \frac{EF_{1veh,i}}{PKT_{1veh,i}} + \sum_j \frac{1}{PKT_{ij}}.\frac{b_{ij}.VKT_{ij}}{\sum_i b_{ij}.VKT_{ij}}.q_j.EF_{1uj} \quad (3)$$

The term $q_j \cdot EF_{1u,j}$ represents the environmental impact of the entire infrastructural network of type $j$ in the perimeter studied.

**Specific infrastructural demand**

One kind of infrastructure can be shared by several modes: this is the case of roads, that are used by taxis, buses, e-scooters, motor scooters, bicycles, etc. Moreover, one mode can use several kinds of infrastructure. This is the case of microvehicles, using cycle lanes, roads, and sometimes sidewalks. The specific infrastructural demand $a_{i,j}$ can vary depending on the type of vehicle $i$ but also on the life cycle stage considered. Several approaches have been developed so far. In ecoinvent transportation LCI, the impact of the infrastructure operation stage is allocated equally between vehicles ($b_{ij}$=1), while the burden from the rest of the life cycle is allocated linearly to the gross vehicle weight (Spielmann et al., 2007). This approach has also been selected for the allocation of road burden to two-wheel vehicles by Leuenberger and Frischknecht (2010). Nevertheless, for roads, Chester developed a different approach, allocating the impact of the maintenance based on damage factors, linear to the axle weight raised to the power of 4 and neglecting the impact of light vehicles, while the rest of the life cycle burden was allocated equally between vehicles. More complex allocations may make more sense from a civil engineering approach, especially a construction stage impact allocated based on the damage ratio of each vehicle. But how to calculate the specific infrastructural demand is a methodological choice.

# 3 Calculation

## 3.1 *Overview of the methodological application*

Figure 1gives an overview of the calculation process developed for the application of the generic LCA method to the case of the FFES breakthrough in Paris. The specificity of the case study comes from the fact that no data were available on the distances traveled by mode in Paris before and after the arrival of FFES. The block on the left is, therefore, an additional survey-specific method, necessary in this case, that will be presented below. When the data are available, equation 1 can be used directly instead. Then, the generic method to applied CLCA to a mobility change on a territory is represented within the second block of the chart. In italics and gray are indicated the elements of the article where to find the assumptions or the results of the calculation in this article.

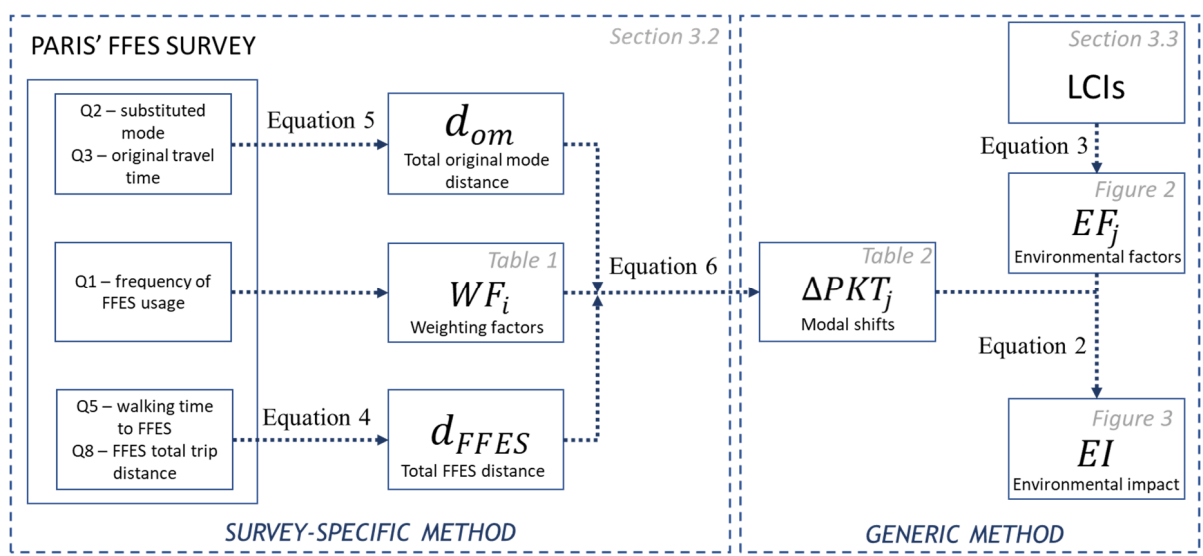

**Figure 1 Overview of the methodology application**

### *3.2 Traffic pattern: modal shift data and assumptions*

#### *3.2.1 Paris survey protocol*

<u>The site</u>

Paris region is divided into Paris inner-city ("Paris intra-muros", 2.1 M inhabitants (IAURIF, 2019)), roughly within a 6-km radius circle, and Paris outer city, divided into two parts: the "inner suburbs", and the "outer suburbs". Most of the FFES are in Paris inner-city, while a few areas in the inner suburbs are also covered. Paris is very well served by public transportation, with sixteen metro lines, five suburban railway lines (RER), four tramway lines, and over sixty-five bus lines (OMNIL, 2018). Shared vehicle fleets have included over recent years: bikes (since 2007), electric cars (2011-2018), and electric motor scooters (since 2016) (small electric motorcycles).

Survey design

The survey has been carried out following a schedule and area sampling matrix (see details in the supplementary material) designed to address both the spatial and time heterogeneities of the FFES usages. Seven locations have been preselected ranging from pure residential areas in the inner suburbs (Issy-les-Moulineaux) to Paris Central Business District (La Defense), also including two leisure sites (the banks of the Seine and the quays of St Martin canal), two major central public transportation hubs (Châtelet and République), and the area of Station F, the biggest startup campus in Paris. Varied time slots and days for the interviews have been selected to capture different user profiles.

Questionnaire and administering

A fifteen minutes-long questionnaire has been developed to address different issues. In this paper, responses to the following questions will be used:

- Q1: How often do you use the ES?

- Q2: If you had no access to an ES, which mode would you have used primarily for this trip?

- Q3: If you had used this other mode of transportation, how long would it have taken you to reach your destination on the same route, in minutes?

- Q4: Where were you going?

- Q5: During this last trip, how long did you walk to reach the e-scooter, in minutes?

- Q6: With which transportation mode did you combine the use of an ES during your last trip using it (several answers possible)?

- Q7: Did you take the scooter with you onboard this mode (s) of transportation?

- Q8: What was the approximate distance of this trip (one-way, in km)?

- Q9: What was the approximate duration of this trip (one-way, in minutes)?

- Plus several socio-economic characteristics questions: gender, age, household situation, income, education

The sampling method was made according to the rule of questioning the first person arriving at a fixed point after the completion of the previous interview. The interviewers provided a short explanation of the research objectives and no other incentive was offered to the respondents. With a response rate of around 85%, we only registered responses from people having already used ES at least once in the past. After the removal of people using exclusively personal ES, our sample counts n=445 responses.

### 3.2.2 New traffic patterns

We want to calculate $= \Delta PKT_i$, and we do not have data directly giving the traffic for each mode of transportation before and after the FFES breakthrough. The modal shifts calculated from our survey based on the last FFES trips are reported in the supplementary material: they mostly come from walking (35%), the metro (23%), and the bus (12%). However, these modal shifts are trip-based: they do not consider the length of the trips. Moreover, they are unweighted regarding the ES usage frequency of each respondent. We will now explain how to transform these figures in weighted kilometer-based data to properly perform the final environmental calculation.

<ins>FFES last trip length</ins>

The distance traveled using FFES during the last trip involving an e-scooter, $d_{j=FFES}$, can be calculated using the following equation 4, with $d_{lt,FFES}$ the total distance of the last trip given by question Q8, $t_{walk}$ the walking time to access the ES given by Q5, and $v_{walk}$ the walking speed, considered equal to 4.7 km/h (based on simulations made using Google Maps on twenty itineraries with different topographies in Paris):

$$d_{j=FFES} = d_{lt,FFES} - v_{walk} \times t_{walk} \qquad (4)$$

Original transportation mode characteristics

Answers to question Q2 and Q3 give respectively the alternative mode and duration of the last trip made using an FFES if no microvehicle had been available. The multimodality of the trips substituted was not investigated further in our survey. Considering that most of the last FFES trips were leisure trips and that only 10% of them were related to work or school, these trips must not concern a large part of complex commuting trips requiring two modes or more. Thus, trips that were made by cars (taxi and ride-hailing, or personal cars), on foot, or by personal two-wheelers are considered as single-mode journeys. For the other modes, the trips are considered as combined with walking. Assumptions are made on the distances to be walked to access and leave each of the shared modes other than the FFES, based on expertise (see supplementary material). Commercial speeds in Paris and the source of these figures are also reported in the supplementary material.

Thus, the distance that would have been traveled using the original mode during the last ES trip if no FFES had been available, $d_{om}$, can be calculated using equation 5, with $t_{lt,om}$ the duration of the last trip (including walking access) if no FFES had been available given by question Q3, $d'_{walk}$ the total distance walked that we assume during the trip (see supplementary material), $v_{walk}$ the walking speed, and $v_{om}$ the original mode average/commercial speed:

$$t_{lt,om} = \frac{d'_{walk}}{v_{walk}} + \frac{d_{om}}{v_{om}} \Leftrightarrow d_{om} = v_{om} \times (t_{lt,om} - \frac{d'_{walk}}{v_{walk}}) \qquad (5)$$

The calculation of the traveled distance using the original mode calculated using equation 5 considers average modal speeds and walking access distances. It is based on approximate individual declarations of time to access the FFES and durations of the trips. These uncertainties lead to unrealistic negative trip durations in 7% of the cases. An analysis of these cases shows that the declared trip duration using the original mode seems too low to be consistent. To correct these outliers, we thus chose to keep the total length of the last trip using FFES instead.

Respondents weighting factor

The last trip distances traveled by FFES and the last trip distances that would have been traveled with the original transportation mode have now been estimated. Nevertheless, we want to obtain the total difference of kilometers traveled under the penetration of FFES, per mode, over one year. Each respondent gave the principal transportation mode he or she would have used if no FFES had been available. But some respondents are heavy users while others only used FFES once: each answer needs to be weighted based on each respondent's FFES usage intensity to get a representative difference in mobility consumption from the sample. To do so, and considering the last trip is representative of the annual behavior, the annual weighting factors $WF$ are indicated in Table 1.

**Table 1 Equivalence between the FFES usage frequency range declared and an average annual number of FFES ride**

| Declaration of FFES usage frequency range in the survey | Annual number of FFES rides | Explanation |
|---|---|---|
| "more than 5 times a week" | 312 | 6 times a week, every week of the year |
| "4 to 5 times a week" | 234 | 4.5 times multiplied by 52 weeks |
| "two to three times a week" | 130 | 2.5 times a week multiplied by 52 weeks |
| "once a week" | 52 | |

| "Less than once a week" | 15 | |
|---|---|---|
| "I only used ES once" or "I stopped using the ES" | Not considered | |

Generalization of the sample to the Paris usages

In Spring 2019, the pioneering and leading operator of FFES in Paris declared a total of 950 000 users in the city (6t, 2019). 1/3 of the users have multi-memberships to increase their chance to get an FFES available (6t, 2019), and they are very likely to be a member of the leading company. Based on these figures, we will consider that one million people have used FFES at least once in Paris over the first year of penetration. Considering our survey sample as representative of the FFES users in Paris, we can calculate $\Delta PKT_j$, the modal shift in Paris mobility consumption generated by the FFES penetration for the mean of transportation $j$, using the equation 6, with $N_{FFES\ users}$ the total number of FFES users in Paris, $n$ our final calculation survey sample, $WF_i$ the FFES usage intensity weighting factor for the respondent $i$, $d_{FFES,i}$ the last trip distance using FFES for the respondent $i$, $d_{j,i}$ (when $j \neq FFES$ index) the last trip distance if using the original mode of transportation $j$ for the respondent $i$, and $\delta_j^{FFES}$ the Kronecker delta function which equals to 1 when j = FFES (=1 based on Table 2 modal ranking) and to -1 in the other case.

$$\Delta PKT_j = \frac{N_{FFES\ users}}{n} \sum_i -(-1)^{\delta_j^{FFES}} d_{j,i}.WF_i \qquad (6)$$

The database has been cleaned following one criterion: all the responses leading to an average speed of the last trip using an FFES higher than 30 km/h have been deleted, considering these trips were probably made using personal ES, as maximum speeds of FFES available in Paris are limited (Maçon, 2019). The new sample size is $n$ =411.

### 3.3    Quantification of the carbon footprint by mode using CLCA

We want to know the environmental impact of moving about in Paris by twelve means of transportation: FFES, walking, personal bicycle, shared bicycle, personal two-wheelers, shared two-wheelers, personal cars, taxi and ride-hailing, bus, metro, RER, and streetcar. RER (Réseau Express Régional, or Regional Express Network) refers to the rapid transit system which serves Paris and its surrounding suburbs.

#### 3.3.1   Methodological choices

<u>Overview</u>

We will perform a process-based CLCA using OpenLCA 1.8.0. The choice of a consequential Life Cycle Inventories (LCI) background dataset - the ecoinvent V3.2 "Substitution, consequential, long-term" dataset - reduces errors in our CLCA results (Weidema, 2017). We have focused on climate change effects, calculated using the CML-IA baseline characterization method (2016) developed by the Institute of Environmental Sciences - Leiden University, with a time horizon of 100 years. This method considers the IPCC 2013 report's global warming potentials. The boundaries of each transportation mode system encompass both infrastructure and vehicle and/or equipment to move, as well as the entire life cycle: stages of production, use, maintenance, and End-of-Life (EoL). The environmental factor of the mode $i$ per pkt $EF_i$ can be calculated using equation 3.

When ecoinvent processes are used in the modeling of vehicle production, we will consider if the vehicle is bought on a global market or not. More precisely, global market processes will be chosen in the case of vehicles as consumer goods, namely personal vehicles but also vehicles provided by private operators. In the case of public transportation vehicles, the choice of the fleet is not only financial but also driven by political and other aspects. Consequently, we will consider the most local markets for France.

<u>The case of active modes</u>

We can classify the twelve Parisian modes in two categories of transportation: active and non-active modes. Only walking and cycling are active modes. The two different kinds of life cycles are illustrated in the supplementary material. The main difference occurs in the use stage, where the motion energy comes from food in the case of active modes, and from fuel or electricity for non-active modes. Choosing an active mode instead of a non-active one implies for the traveler extra output energy. There is no consensus about whether considering this energy in LCA. For instance, Dave (2010) considered that commuters do not change their food consumption either by walking "given a reasonable walking distance and assuming sufficient nutrition is available", or by cycling. But Thorpe (2016) considered the extra food consumed by cycling instead of driving, and its related GHG emissions depending on the type of diet of the person. This second approach is interesting to study the impact of specific diets on physical activity impact. Nevertheless, more than half the population in Europe (WHO, 2019) and the United States (CDC, 2019) do not meet the World Health Organization recommendations on healthy physical activity level (WHO, 2010): we, therefore, chose not to consider the extra food consumed under active mobility, given the fact people should be generally more active. We also did not consider the environmental impact of shoes for walking, due to their multifunctionality and probable low impact. For cycling, we considered the manufacture of an average bicycle on the global market and its maintenance (Leuenberger and Frischknecht, 2010).

### 3.3.2 *Free-floating e-scooter Life Cycle Inventories*

LCI come from an LCA model developed for personal ES in the city of Paris (de Bortoli et al., 2020c) and are detailed in the supplementary material. This model considers a first-generation ES with an autonomy range of around 20 km in Paris real traffic conditions and a 0.335 kWh battery capacity. The production, transportation, and EoL stages are supposed similar for a personal and a shared ES. On the other hand, the use and maintenance stages will be different. The life cycle mileage of the ES, i.e. the number of kilometers traveled before breaking down, as well as the spare parts to be replaced and the maintenance operations, vary due to differences in the way personal and shared ES are handled. Indeed, rough usages or even vandalism are often reported by free-floating microvehicle operators. A charging operation requiring transportation also occurs in an FFES life cycle, while it is not the case for personal ES. As a base case, we considered a 500 charging cycle lifespan for the battery as specified by the manufacturer, and a 3 750 km lifespan, i.e. 1 year of total lifetime if used 11 km per day (average scenario in the study by Hollingsworth et al. (2019)). The ES servicing baseline scenario to charge batteries is considered using light commercial vehicles (LCV) that come from and take the FFES to a suburban warehouse, 20 km outside Paris, and travel 10 km in Paris. Each LCV carries 100 ES, for a return trip distance of 0.9 km per ES every day to charge and distribute them. We consider the production, maintenance, and EoL treatment of the LCV based on ecoinvent processes, and a lifespan of 150 000 km. The LCV consumption is 0.16 L/km and the tail-pipe carbon emission 0.403 kgCO2/km (ADEME, 2019).

### 3.3.3 *Conventional original vehicles LCI*

The impact of the vehicle life cycle, not including the use stage, is calculated based on ecoinvent as synthesized in the supplementary material. We present the main assumptions and characteristics of the model. The weights expressed in tons refer to the metric ton.

The average car in France is 8.2 years-old with a 103 771 km mileage (Kolli, 2012), for an annual mileage of 12 650 km. Considering a car can be operated for 2x8.2=16.4 years, its average lifetime mileage could be estimated at around 200 000 km. Following the French statistics (Compte, 2018), we considered a fleet of personal cars made of 1278-kg vehicles, 61% of them using diesel, and 39% gasoline. Consistently with most car LCA studies, we considered a mileage of 150 000 km over the life cycle. For taxis and ride-hailing, we considered a fleet composed of 82% diesel cars, 7% gasoline cars and 11% electric cars, representative of this market in the Parisian region (Ducamp and Tanca, 2019), a heavier weight of 1400 kg (+250 kg of electric battery if relevant) representing an average French sedan car, and the same lifetime mileage.

The bicycles, electric or not, are manufactured in China and transported to Europe. A personal bicycle is supposed to weigh 17 kg like the ecoinvent reference, while the shared bicycles in Paris weigh 20.6 kg, with an additional electric motor of 2.2. kg and a lithium-ion battery of 3.9 kg on the electric version (Smoovengo, 2017). The lifetime mileage is assumed to be 15 000 km for a personal bicycle (Leuenberger and Frischknecht, 2010) and 4 500 km for the station-based bicycles, similar to the worst-case scenario assumption in Luo et al. (2019) due to vandalism and consistently with the lifespan of the electric bicycle battery according to ecoinvent assumptions.

The shared two-wheeler is a battery electric motor scooter of 120 kg with a 2.1 kWh fixed battery and a 1.9 kWh switchable battery (Fontanier, 2017), for a global battery weight estimated around 8 kg. We considered the personal two-wheeler to be an internal combustion engine (ICE) motor scooter weighing 180 kg. Both shared and personal two-wheelers are assumed to have a 50 000 km mileage (Leuenberger and Frischknecht, 2010).

The metro train was assessed by scaling down the ecoinvent process of regional train manufacture. The ecoinvent train weighs 171 tons while a 5-carriage metro weighs around 131 tons. The RER train is scaled up based on an average mass of 230 tons. The streetcar is scaled up based on the ecoinvent process, a Parisian tram vehicle weighing 52 tons while the model in ecoinvent weighs 21 t. The bus manufacturing, maintenance, and EoL are approximated directly by the related ecoinvent process for a standard 13m-long bus, as Paris' bus fleet is composed of 80% standard buses, the rest being a mix of double buses and minibuses (OMNIL, 2019a). The theoretical mileages of the four previous vehicles come from Spielman et al. (2007), except for the mileage of buses that comes from Parisian data (ADEME et al., 2018), and are specified in the supplementary material.

Assumptions for the use stage of these vehicles are now presented. Occupancy, consumption, and exhaust emissions are also synthesized in the supplementary material. Two original shared modes require servicing activities: the docked bicycle and the free-floating motor-scooter. We did not consider any servicing for the shared motor scooters, due to a lack of data. Nevertheless, the leading Parisian operator uses 100% electric vehicles for scooter maintenance and battery charging (Cityscoot, 2018). Using the French Impact-ADEME database, we calculated that a French electric car emits in average 74 gCO$_2$eq/km (two-thirds of this impact being due to car manufacturing), i.e. 63% less than a standard car. Moreover, batteries are swappable and fit in a small volume. These two characteristics must reduce considerably the environmental impact of servicing. As the shared motor scooter trip substitution is very secondary, the final impact will not be affected by this data gap (a maximum extra emission retro-calculated equals to 0.7% under the worst-case scenario, i.e. consisting of doubling the impact when including servicing). The docked bicycles need to be rebalanced between the stations: we considered a rebalancing distance of 27.5 m/bike-km, similar to the base scenario assumption in Luo et al. (2019), made by Euro 5 trucks (3.5 to 7.5-ton capacity).

The characteristics of the vehicle energy consumption during the use stage are detailed in the supplementary material. We considered data as close as possible to the Parisian context. We assume that all the electric vehicles are fed with high-voltage electricity except the e-scooters. For public transportation, the main operator in Paris, RATP, gives the GHG emission of metros, streetcars, and RER considering occupancy rates and electricity consumptions from the grid for 2017, and an emission factor of 48 $gCO_2eq/kWh$ for the electricity. We can retro-calculate the consumption per passenger-kilometer based on these data, to calculate the carbon footprint for the use stage with a consequential approach instead of an attributional approach, and with a background dataset consistent with the rest of the model (SNCF, 2019). Metro and RER are powered with high-voltage electricity, from the French electricity market. Their occupancy is estimated with traffic data for the Paris metropolitan area (OMNIL, 2019b, 2019c), thus underestimated for the inner-city. The tail-pipe GHG emissions of gas-powered cars are taken from the French ADEME database that relies on HBEFA models (Matzer et al., 2019), and modeled using an elementary flow of carbon dioxide from fossil sources emitted in a high-density area, to which we added indirect emissions due to the supply chain (ADEME, 2019). The gas is low-sulfur, with a density of 820 $kg/m^3$ for diesel and 720 $kg/m^3$ for gasoline. The consumption of the electric taxis is estimated at around 25 kWh/100km. Taxi and ride-hailing cruise empty 50% of the time according to consultation of professionals (ADEME, 2019), and with 1.7 pax/vehicle the rest of the time based on a Parisian survey (6-t, 2015), leading to an average occupancy of 0.85 pax/vehicle. The direct emissions of buses per passenger-kilometer in Paris come from SNCF for the year 2017 (2019), the average number of people in a 12 meter-long bus is calculated based on Parisian bus data (OMNIL, 2019c, 2019b) and the average consumption for indirect emissions taken as being equal to 0.35 kg per vkt (Spielmann et al., 2007).

### 3.3.4 Infrastructure inventories, demand factor, and allocation factor

To calculate the environmental burden from the life cycle impact of the infrastructure $EF_{infra,i}$ on each transportation mode $i$, we used equation 3. In the case of an infrastructure type only used by one kind of vehicle, e.g. the metro tracks, the allocation is directly based on the demand factor, i.e. the inverse of the PKT on the infrastructure. This is also the case for a kind of infrastructure supporting vehicles with the same allocation factor and occupancy, e.g. cycle lanes that cater for bicycles and ES in this study.

First, we calculated $q_j.EF_{1u,j}$, i.e. the environmental impact of the Parisian infrastructural networks affected by the breakthrough of FFES: the metro railway, the RER railway, the streetcar railway, the roadway, the sidewalks, and the cycle lanes. Each of these categories presents physical heterogeneities that are difficult to catch without proper asset management databases. The unitary impact of each type of infrastructure on a life cycle approach, $EF_{1u,j}$, was modeled using field data when possible, as well as generic data and ecoinvent (see details in the supplementary material). Because no specific data are available, the unitary LCI for the streetcar infrastructure directly comes from the ecoinvent process "Tram track construction, CH", that models the impact of one meter of the infrastructure over one year based on the entire life cycle of a 6-meter wide double track concrete section, considering construction, renewal, operation, maintenance, and decommissioning. Some elements are considered out of the system boundaries, namely all the operations that happened only at the first construction stage or subsystems that will not be decommissioned, consistently with a consequential approach. This is the case, for instance, of the metro and RER tunnels, or the roadway earthwork. Thus, the ecoinvent process "railway track construction, CH" was used as a proxy to model metro and RER infrastructure LCI for one meter and one year, as no data are available for these types of infrastructure in Paris. We did not consider coefficient ponderation based on width as most of the impacts may come from the rail (de Bortoli et al., 2020a) and the operation. The other unitary LCI are developed for the case study. The impact of pavement construction on climate change mainly comes from materials and manufacturing, then from transportation, while the impact of building machines is very low (Cuenoud, 2011; Kucukvar et al., 2014; Tatari et al., 2012; Vidal et al., 2013). This latter was not considered. Transportation over 50 km of the gravel and bitumen will be considered. A French process for the manufacture of Hot Mix

Asphalt (HMA) is proposed (see supplementary material) based on field data. The type and quantity of energy consumed in asphalt plants for bitumen heating, gravel drying, and HMA mixing come from Eurovia, a company that owns 35% of the national plants (de Bortoli, 2018). Water consumptions and emissions to water and air come from a survey made on 8 asphalt plants in France (USIRF, 2016). The cycle lane network in Paris measures 742.1 km (Ville de Paris, 2016) for a surface of 106 hectares (Breteau, 2016), i.e. is 143-cm width. It is delimited from the pavement and the sidewalk using two curbs of type T1 (54 kg of pre-casted concrete per linear meter according to manufacturers' catalog). Its structure is modeled as a 15-cm layer of gravel covered with a 2.5 cm layer of HMA (Conseil général des Yvelines, 2011) made of 6% bitumen and 94% gravel. The typical lifespans of the rolling course and the base are estimated to be namely 20 years and 60 years. For the sidewalks, we considered a lifespan of 20 years and a structure made of a subbase layer of 15 cm of gravel covered with 2.5 cm of mastic asphalt (10% bitumen and 90% gravels). Roads were modeled as a French T3 type of road (Corté et al., 1998), carrying up to 150 heavy vehicles per day and per direction, made of 15 cm gravel for the pavement subbase, 11 cm bitumen-bound graded aggregate (with 4% bitumen and the rest made of aggregates) for the base, and 4 cm HMA for the rolling course (with 6% bitumen). The lifespan of these layers is respectively 60, 30, and 15 years.

The volume of each network and supported mobility in Paris is synthesized in the supplementary material and comes from GIS treatments for the surfaces of sidewalks, cycle lanes, and parking lots, roads and parking lots, as well as bus lanes (Breteau, 2016), and from the City of Paris for the length of the metro railways (Ville de Paris, 2016), approximated as if the entire network was in the inner city. The RER railway length is estimated at around 80 km, i.e. the length of the underground sub-network, while the streetcar tracks length measured on a map gives 29 km. The demand and supply per mode in Paris - $PKT_{i,j}$, and $VKT_{ij}$ - is specified in the supplementary material. Public transportation data come from regional surveys for 2018 (OMNIL, 2019b, 2019c). For the other modes, the source of the data is the latest French transportation survey delimited to Paris inhabitants (DRIEA et al., 2013) corrected by the coefficient 2.08 to add traffic due to visitors. This coefficient has been calculated by comparing 14.6 billion, the total number of trips made in Paris every year, to 7.03 billion, the number of trips made by Paris' inhabitants in the inner-city. We need to estimate the vehicle-kilometers made by taxis and ride-hailing cars in one year. One taxi travels 57 700 km a year (CGDD, 2018). We assumed a ride-hailing car travels 25 000 km a year as many drivers deliver this service as a part-time job. Paris' prefecture counts almost 17 500 taxis in the city (CGDD, 2018), a figure close to the number of ride-hailing cars in the city estimated at 45% of the total of taxis and ride-hailing cars (CGDD, 2018), i.e. around 14 000. These figures lead to 1.36 billion vehicle-km traveled a year, and given that 4.85 billion of car-kilometers are traveled in Paris inner-city every year, this leads to personal car traffic of 3.49 billion car-kilometers a year.

Finally, we calculated $b_{ij}$, to attribute a share of the environmental burden from the infrastructure to each mode. In the case of shared infrastructure – only the pavement is affected here – the relative use of the infrastructure for each mode must be calculated through an allocation factor. If we consider that road pavements are designed based on axle weights over the lifespan, only heavy vehicles are responsible for these impacts (Chester, 2008). Nevertheless, minimal mechanistic resistance is required to support light traffic as well. We hence considered that a heavy vehicle and a light vehicle are responsible for the same infrastructure burden. Bicycles and ES can run on cycle lanes as well as on pavements with other kinds of vehicles. We did not allocate any of the pavement impact to them, considering pavements are designed for heavier vehicles. The figures about vkt on the Parisian roads by non-microvehicles come from the last data surveyed in the city in 2014 (AIRPARIF, 2018).

Paris's inner-city traffic and final allocation factors in 2017 and 2018 are detailed in the supplementary material.

# 4 Results and interpretation

## 4.1 Consequences of the FFES disruption on Paris mobility

The annual mobility consumption changes in Paris, generated by the breakthrough of FFES over one annum, are calculated with equation 6. We present the results in Table 2. Under our hypothesis and despite the 5% of induced trips, the mobility consumption has decreased by 38.5% on a kilometer basis with FFES, i.e. a saving of 150 million kilometers traveled under a one million users assumption. We also recall the unweighted trip-based modal shifts calculated directly from our survey in the last column of Table 2. The substantial differences between the weighted kilometer-based and the unweighted trip-based modal shifts highlight the importance of two treatments of the raw survey data: (1) the weighting of each response according to the frequency of use of the FFES as well as (2) the transformation of trips into traveled distances by mode. These stages have very important effects on the understanding of mobility consumption intensity. Different explanations can be given. First, average trip distances vary significantly between transportation modes: walking trips are for instance relatively short compared to motorized trips. Secondly, the network density determines traveled distances between points A and B. Transit vehicles run on specific routes, not necessarily the shortest possible. These constraints create extra distances to travel from A to B, compared to FFES that allows switching to more direct routes, only constrained by the existence of ridable streets. Table 2 shows that two-thirds of the weighted modal shifts come from public transportation: 50% from the metro, 9% from the RER, and 6% from the bus. The modal shift from streetcars is negligible. Modal shifts from cars, either taxi, and ride-hailing or personal automobiles, accounts for 7% of the kilometers shifted. Bicycles, either shared or personal, accounts for 9%, and walking for 13%.

**Table 2 Mobility consumption changes generated by the breakthrough of FFES at the scale of Paris over one year**

| | Changes in the kilometers traveled | | | Weighted kilometer-based modal shifts (%) | Unweighted trip-based modal shifts (%) |
|---|---|---|---|---|---|
| | Survey sample | Paris base case (1M users) | Paris variant (0.5M users) | | |
| Equation 6's term | $\sum_i -(-1)^{\delta_j^{FFES}} d_{j,i}. WF_i$ | $\Delta PKT_j$ | $\Delta PKT_j$ | | |
| FFES | 9.73E+04 | 2.37E+08 | 1.18E+08 | | |
| Walk | -2.09E+04 | -5.09E+07 | -2.54E+07 | 13.2 | 34.6 |
| Personal bicycle | -8.30E+03 | -2.02E+07 | -1.01E+07 | 5.2 | 3.9 |
| Shared bicycle | -6.33E+03 | -1.54E+07 | -7.70E+06 | 4.0 | 3.5 |
| Electric motor scooter | -4.46E+03 | -1.09E+07 | -5.43E+06 | 2.8 | 1.5 |
| Personal motor scooter | -3.30E+03 | -8.02E+06 | -4.01E+06 | 2.1 | 0.2 |
| Car | -7.35E+03 | -1.79E+07 | -8.94E+06 | 4.6 | 3.7 |
| Ride-hailing | -3.37E+03 | -8.19E+06 | -4.10E+06 | 2.1 | |
| Taxi | -5.53E+02 | -1.34E+06 | -6.72E+05 | 0.3 | 6.1 |
| Bus | -9.68E+03 | -2.36E+07 | -1.18E+07 | 6.1 | 11.5 |
| Metro | -7.88E+04 | -1.92E+08 | -9.59E+07 | 49.8 | 23.3 |
| RER | -1.48E+04 | -3.60E+07 | -1.80E+07 | 9.3 | 2.2 |
| Streetcar | -4.04E+02 | -9.84E+05 | -4.92E+05 | 0.3 | 0.4 |
| *Total automobile shift* | *-1.13E+04* | *-2.74E+07* | *-1.37E+07* | *7.1* | *9.8* |
| *Total public transportation shift* | *-1.04E+05* | *-2.52E+08* | *-1.26E+08* | *65.5* | *47* |
| Total shift | -1.58E+05 | -3.85E+08 | -1.93E+08 | 100.0 | 100 |
| **Change in total distance traveled (km)** | **-6.10E+04** | **-1.48E+08** | **-7.42E+07** | **-38.5** | |

## 4.2   *Life cycle carbon footprint for most of the modes of transportation in Paris*

Details of the calculation results from the application of the method described in part 2 can be found in the supplementary materials. The estimates of the GHG emissions per transportation mode in Paris are presented in Figure 2. They highlight the low contribution of the infrastructure in a city with very dense traffic, except for the streetcar mode where it represents 54% of the emissions. For all the modes powered by electricity, the impact of the use stage in absolute value is very low, but it contributes to respectively 80% and 58% of the total emissions for the RER and the metro modes. Taxi and ride-hailing is the most emitting mode with 299 gCO$_2$eq/pkt, due to low occupancy as well as large and heavy vehicles that consume thus emit more than the average personal car, presenting an emission of 209 gCO$_2$eq/pkt. Private motor scooters emit 135 gCO$_2$eq/pkt, which is quite similar to the bus, the less environmentally friendly mode among the public transportation options in Paris, with 133 gCO$_2$eq/pkt, due to a rather low occupancy (17 passengers per vehicle) and a modeled fleet still using diesel. Shared bicycles emit around 59 gCO$_2$eq/pkt due to the burden of the vehicle manufacture stage. Indeed, the rather short lifetime mileage of these bicycles in Paris is a major environmental drawback. Nevertheless, it is still twice better than the shared e-scooters, that emit 109 gCO$_2$eq/pkt, half of the emissions coming from the servicing stage, and the other half from the manufacture also due to short lifetime mileage. The less emitting modes are shared motor scooters, streetcars, the RER, the metro, and walking, respectively emitting 28, 20, 9, 8, and 2 gCO$_2$eq/pkt.

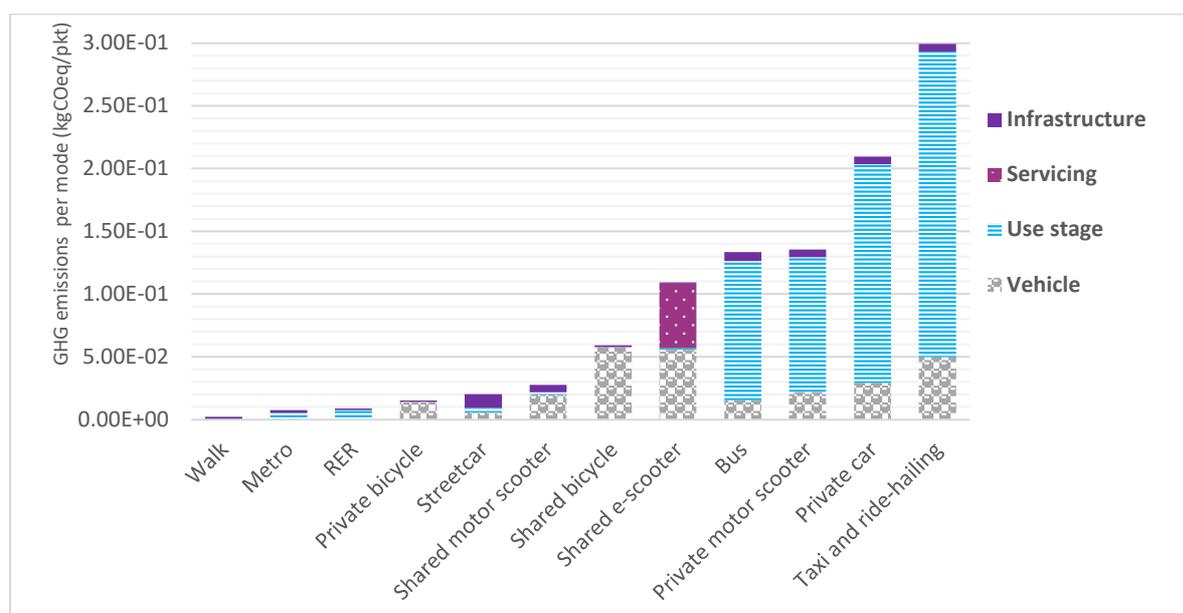

**Figure 2**　　　**Life cycle carbon footprint of the main modes of transportation in Paris**

### 4.3 *Marginal effects of the FFES breakthrough on GHG emissions in Paris over one year*

Calculations show that the breakthrough of FFES raised the GHG emissions of the mobility sector in Paris by 12.9 thousand tons of $CO_2$eq over one year if one million people were FFES users. Under a 500 000 users assumption, this loss is cut by a factor of two. Details of the marginal emissions over one year are presented in Figure 3. Most of the gains are brought by the modal shifts from buses, private cars, taxis, ride-hailing, and private motor scooters. Most of the distance of the trips substituted by the FFES was traveled by metro as showed in Table 2. As this is a very low emission mode in Paris, the gain is far less than the losses due to the FFES. Figure 3 also highlights the feeble importance of the infrastructure part (3% of total marginal emissions), thus of its modeling and especially its allocation to vehicles. This is specifically due to the intense use of Paris's transportation infrastructure. Most of the marginal emissions come from the use stage (57%) and the rest of the vehicle life cycle (40%). To be climate-positive in Paris under our baseline assumptions, the FFES must emit less than 56 $gCO_2$eq/pkt.

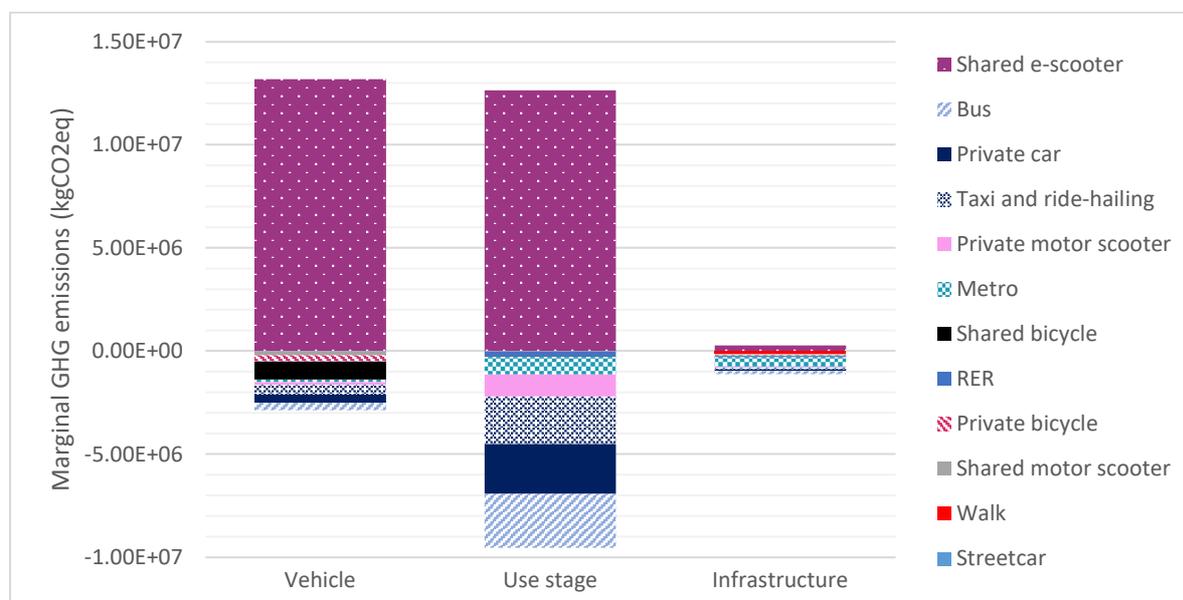

**Figure 3 Marginal GHG emissions due to the FFES disruption, per mode, over one year for Paris' mobility, with detailed contributions from the infrastructure, the use stage, and the vehicles**

### 4.4    Scenario and sensitivity analysis

#### 4.4.1    Selection of the influential parameters

In absolute value, two-thirds of the marginal emissions shown in Figure 3 are due to two major contributions: the FFES manufacturing (33% of the marginal emissions) and the FFES servicing (31% of the marginal emissions). Moreover, the low carbon intensity of the French electricity mix is partly responsible for the high environmental performance of the metro, the RER, and the streetcar, accounting for 59% of the kilometers shifted. Thus, a higher electricity carbon intensity would change the consequential carbon footprint of the FFES penetration. To consider the impact of some variability on these three influential factors, we performed a one-at-a-time sensitivity analysis on the FFES lifetime mileage, the FFES servicing scenario, and the electricity mix.

#### 4.4.2    FFES lifetime mileage

The lifetime mileage is frequently brought under the spotlight when it comes to the FFES environmental performance. A range of [300;15000] km lifetime mileage was selected based on a worst-case scenario (Hollingsworth et al., 2019; Quartz, 2019) and the best-case scenario observed on ES users forums. The results are presented in Figure 4. It shows that, in the specific Parisian context, under our baseline servicing scenario, with the current mobility system and the first observed modal shifts brought by the FFES breakthrough, this disruption is an environmental loss with regards to climate change whatever the FFES lifetime mileage, going from an extra 167 thousand tons of $CO_2eq$ to an extra 3 thousand tons, respectively for a 300 km and a 15000 km lifespan.

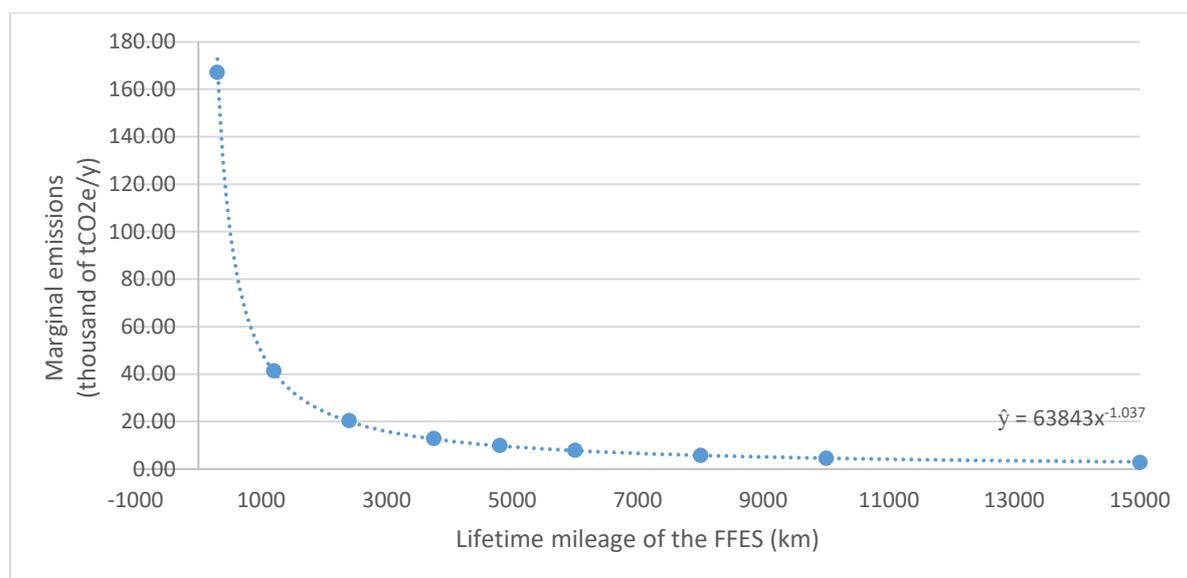

**Figure 4  Consequential climate change contribution of FFES in Paris over one year depending on the lifetime mileage scenario of the FFES (regression shown in dotted line, with the equation and R-squared indicated on the right)**

### 4.4.3    FFES servicing scenarios

This second parameter was chosen based on the observation of Hollingsworth et al. (2019), validated by our results in Figure 2 and Figure 3: half of the impact coming from using an FFES is generated by its servicing, assuming heavy gas-powered vehicles moving FFES over several kilometers every day. In Figure 5, we compare the results for our baseline servicing scenario (labeled "LCV 90 km 100 ES") with six alternative servicing scenarios, observed or not in Paris, presented in supplementary material: "LCV 90 km 50 ES" for an LCV charging only fifty ES and traveling 90 km, "juicer 10 km car" for an entrepreneur charging eleven ES in a car traveling 10 km, "swappable battery 90 km car" for one hundred ES batteries charged in a car traveling 90 km, "swappable battery 90 km car" for the same scenario with only 45 km traveled, "riding juicer" for a person charging two ES on another ES over 4 km as a side activity; and "walking juicer" for a person taking two ES on foot over 2 km as a side activity. It shows that no scenario can lead to a positive climate effect of the FFES disruption in Paris under our baseline model's assumptions. If no servicing at all was done for FFES, their carbon balance would still be + 500 tons of CO2eq over one year.

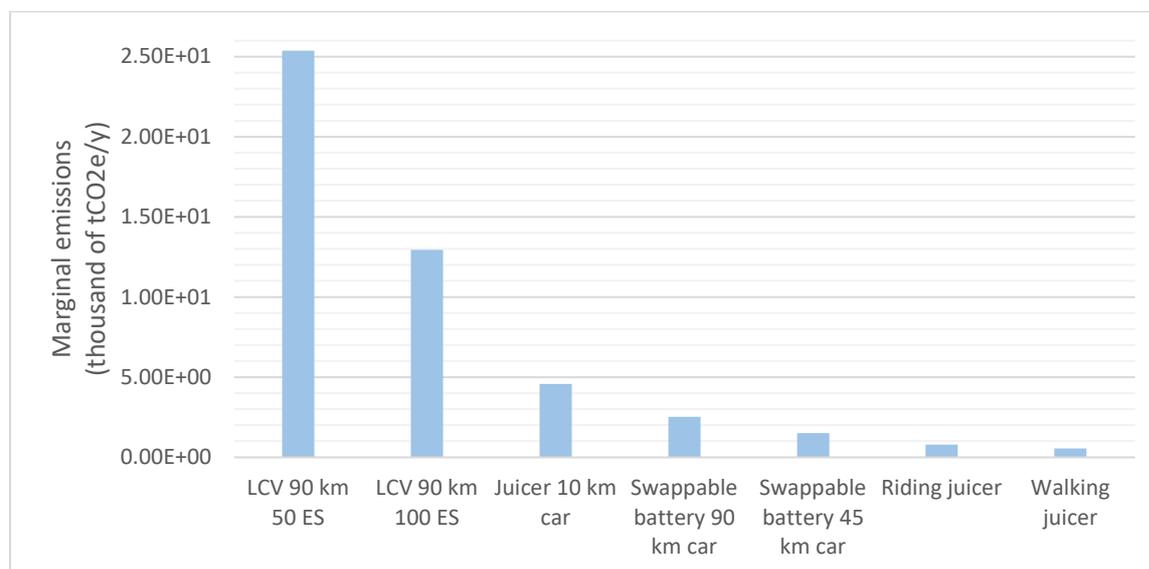

**Figure 5 Consequential climate change contribution of FFES in Paris over one year depending on the FFES servicing scenario**

### 4.4.4   Sensitivity to the electricity mix

This last parameter was chosen for sensitivity analysis because of the specificity of France and Paris. The French electricity mix comes 70% from nuclear plants, with a lower impact on climate change than electricity mixes based on fossil fuels like the case in the United States (average mix), Germany, or China (see supplementary material). We simulated the extra-burden generated by the FFES over one year with eleven alternative electricity mixes for the use stage of the different means of transportation, *ceteris paribus*. In Figure 6, the results show the linearity of the model to the carbon footprint of the electricity mix. The higher the carbon intensity of the mix, the better the FFES impact on the climate. Indeed, the breakthrough of the FFES provokes a decrease in electricity consumption. It is explained by the fact that 60% of the kilometers substituted by the FFES were previously traveled using the metro and the RER (most of the rest of the kilometers substituted rely on fossil fuel). These means of transportation are electric, and respectively consume 7.08 10-2 and 1.13 10-1 kWh/pkt during the use stage (the electricity consumption on the rest of the life cycle has a marginal influence on the environmental results). The FFES only consumes 0.335 kWh / 20 km = 1.68 10-2 kWh/km. Replacing metro and RER trips with FFES trips thus largely decreases the electricity consumed. The higher the carbon footprint of the electricity, the more GHG emissions saved from this electricity consumption savings. Of course, in other countries, mobility systems would be different, and each specific case needs to be simulated precisely with context-specific data on the mobility system and trip substitutions as well as on the local technosphere. But we can conclude that for a highly electrified urban mobility system like in Paris, as FFES are very light vehicles that consume a low quantity of energy compared to all non-micro- electric vehicles, this service can be beneficial to limit climate change. Namely, under the baseline scenario assumptions, the FFES service becomes an asset to the climate change issue when the electricity carbon footprint is higher than 1.04 kgCO$_2$eq/kWh (e.g. for China's electricity mix).

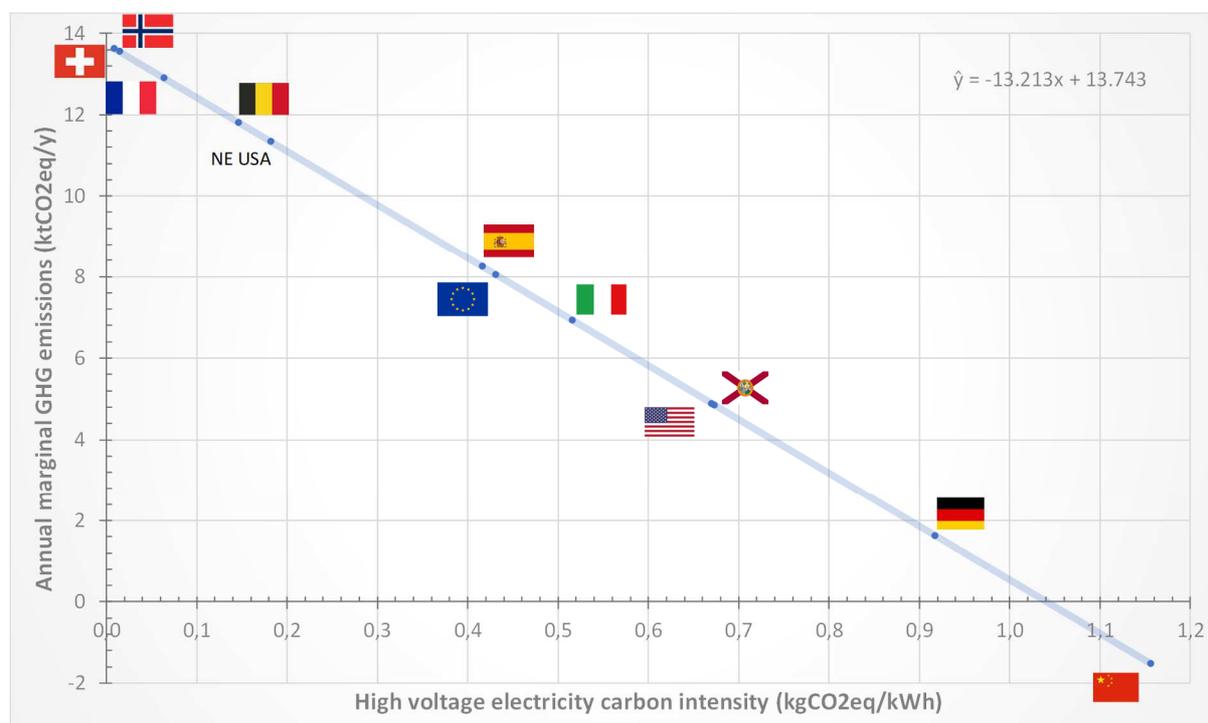

**Figure 6 Extra GHG emissions due to FFES over one year, depending on the carbon intensity of different country's electricity mixes**

# 5   Discussion

## 5.1   *Methodological recommendations*

### 5.1.1   *A life cycle approach to robustly assess current urban mobility systems*

In recent decades, many methods have been developed to monitor and reduce the various transportation environmental impacts triggered by the dominant car-oriented urban planning. But transportation systems have dramatically changed these last 15 years, especially in urban areas with the rise of shared and/or micro- vehicles. The environmental profile of standard fuel-powered means of transportation was globally correctly estimated through the sole consideration of vehicle consumption and emissions. Indeed, in terms of GHG emissions, the use stage accounts for 85 to 90% of the impact of buses, private cars, private motor scooters, or taxi and ride-hailing in Paris (see Figure 2). Nevertheless, this is different either for many new mobility systems or for electrified modes consuming low impact electricity. The paper shows how the GHG emissions of (shared) microvehicles currently come mainly from the manufacture of the vehicles, and potentially the servicing if any. The manufacturing stage contributes from 49% to 97% of the total impact, respectively for FFES and shared bicycles. For non-micro electrified modes, the main contribution can either come from the manufacturing, e.g. for electric cars (70% of the total impact), or the use stage (resp. 58 and 79% for metro and RER). Moreover, non-massive modes using a dedicated infrastructure can present substantial impact contributions from their infrastructure (de Bortoli et al., 2017). This is the case of the streetcar in Paris where the railway contribution from the railway represents 54% of the total impact. Finally, when it comes to GHG emissions, truncating the assessment perimeter of transportation systems to their use stage is now an erroneous approach that must be corrected, by using full life cycle approaches. On other indicators than GHG emissions, the approach was potentially already mistaken for road modes. For instance, human toxicity and ecotoxicities impact contributions mainly come from metal emissions within the vehicle production stage (de Bortoli et al., 2017). This study focuses on GHG emissions, but LCA is a multicriteria method than can prevent mono-criterion biases, i.e. environmental burden shifting from one impact category to another that are not detectable when considering only one criterion. Finally, a life cycle approach is necessary to assess the environmental impact of transportation policies or technologies both to consider all the components – infrastructure and vehicles – and their life cycle stages – from raw material extraction to the EoL – and a full panel of environmental impacts. This only can generate robust results for decision-making, free from biases due to (1) truncated system boundaries and (2) monocriterion environmental assessment. Thus, we recommend a more complete multicriteria assessment of the FFES impact in the future.

### 5.1.2   The choice of CLCA to consider the rebound effects and future changes

Consequential approaches are classically used in transportation policies, and for instance in French socio-economic appraisals of transportation infrastructure, to capture the rebound effects from a change in the mobility system, like shorter traveled distances or modal shifts. Indeed, rebound effects are known to have potentially huge consequences on mobility systems' environmental burden. Nevertheless, full life cycle environmental appraisals in transportation have mainly been conducted using ALCA, which does not consider these effects. To our knowledge, no specific CLCA method for multimodal transportation problems has been developed so far. Our CLCA method should be selected in the future to assess the environmental consequences of any transportation policy, as a consequential approach in LCA is the only method to prevent biased decision-making due to (3) omission of indirect systemic effects (i.e. rebound effects) and (4) future changes that could modify the environmental assessment results.

## 5.2 Uncertainties and sources of errors

### 5.2.1 Accounting for uncertainties

The numerous sources of uncertainties in LCA and how to analyze them was largely discussed in the literature (Bamber et al., 2020). The Monte Carlo simulation is the most popular method (Bamber et al., 2020), despite a certain shadow or confusion on the mathematical ground to apply it to LCA studies, including probability distribution functions and independence of parameters among others. Recently, even the basic number of runs - typically chosen equal to 1000 or 10000 - has been discussed, and its basic use on Pedigree factors deterred (Heijungs, 2020). Because other methods are more complex and time-consuming, no quantitative uncertainty analysis was performed in our study, despite the fact the results would benefit from such a complementary analysis to ensure the robustness of the findings. Nevertheless, we will qualitatively discuss the sources of uncertainties specific to our case study.

### 5.2.2 Limitations due to traffic data and survey

The carbon footprint impact of the FFES penetration was estimated under the assumption of one million FFES users in Paris, with only 15% of unique riders and 1.5% of droppers based on our survey. We made an extrapolation on the number of trips per year per type of user from the number of uses per week declared. It gives an estimation of the number of trips per year, equal to 72 million based on a weighted average trip length of 3.3 km. This number represents 0.5% of the number of trips in the Paris region. It seems consistent with another estimation made by 6-t, within 24 and 59 million trips per year for the only people who both use the FFES in Paris and live there, i.e.one third of total users of FFES in Paris (6t, 2019). In our model, the result is linear to the number of users: if it drops to 500 000, impacts are cut by a factor of two. We did not consider seasonal effects, whereas micromobility behaviors depend on the weather, as shown in the example of shared bikes in New York City (An et al., 2019).

The question of intermodality is also very important. An intermodal trip is a trip made using more than one transportation mode. Rare are the trips not combined with walking, and the threshold to consider them intermodal is not clear. 74% of the trips in our study were combined with walking. The twenty-six other percent were combined with other modes: 26% with the bus, 32% with the metro, 21% with several modes, 5% with taxi or ride-hailing, 4% with a bicycle (shared or personal) and 6% with cars (shared or personal). We excluded these trips from our estimate due to uncertainty on the modal share and thus the modal distances. A specific survey would be necessary to include intermodality data.

Finally, consistency regarding perimeters considered in different datasets – e.g. infrastructural network length or traffic – is an important point. Statistics on broader perimeters than the one in the study will undoubtedly bring uncertainty. For instance, Paris inhabitants' trips data, extracted from the latest French transportation survey (EGT 2010), have been considered. But one-third of their trips have a destination outside Paris inner city. These trips probably have different characteristics than the ones within Paris, bringing some uncertainties.

### 5.2.3 The consequential approach of disruptions

Other uncertainties in the model come from adopting a consequential approach instead of an attributional approach. When it comes to transportation public policies, and especially electromobility penetration, this switch is recommended to capture market effects in prescriptive LCA made to enlighten decision-maker choices and to design regulations. Indeed, electromobility generates higher electricity consumption, lower oil consumption, and modifies vehicle supply chains. All these changes must be considered on a marginal approach. But an LCA can be consequential to different extents and in different dimensions touching different phases of the assessment. Our goal and scope are consequential (phase 1), and we chose consequential LCI (phase 2). Some uncertainty is due to the background dataset. The consequential LCI provided up to ecoinvent 3.3 would have several shortcomings, and, concerning electricity mixes, they could lead to unrealistic marginal mixes in several countries (Vandepaer et al., 2019). The use of energy scenarios instead of ecoinvent processes would allow the evolution of the electricity system to be considered within the definition of the marginal mixes. Moreover, the consequential phenomena considered are always limited due to a lack of data. Especially, the LCI to assess the original transportation modes need further investigations based on specific Parisian infrastructure maintenance data that the operators were not able to provide yet: which maintenance practices, and how they are impacted by a change in mobility consumption. Maintenance models are probably non-linear to time as aging models are often not, and traffic demand is not linear either. Relatedly, this consequential assessment is based on a general equilibrium hypothesis. We considered that the reduction in passenger-kilometers traveled had some environmental impacts. Nevertheless, this reduction only has a real impact on the condition that the public transportation offer is adapted to the new demand. If the modal shift only empties buses, streetcars, metros, and RER, without changing the number of vehicle-kilometers traveled, there is an effect on comfort, especially in rush hours, but no environmental impact. Not to mention a possible rebound effect, leading to more mobility. A more complex model could be developed with more sophisticated data. As CLCA is supposed to catch changes in production capacities (Earles and Halog, 2011), a longer time horizon than one year should also be considered. Nevertheless, changes in ES production have already occurred, and this preliminary study is based on early traffic change evidence.

As FFES is a new phenomenon, the LCI of the ES bring many uncertainties. ES designs are evolving to make stronger microvehicles with higher lifetime mileage. Types of materials are changing too, as well as quantities of materials, battery range, lifespan, maintenance, and EoL management. From the basic Xiaomi M365 model that was not designed for shared usages, most companies have now switched to more robust designs, sometimes with higher autonomy ranges. Dependence between the ES design - thus material quantities – and its life cycle mileage is very likely, but no data are available to analyze this causality. Some operators have now maintenance workshops where they can reuse up to 95% of the spare parts from out-of-order FFES (Lelievre, 2019). How many times these parts can be reused and for which final mileage would be a piece of information required to perform a more robust LCA. The question of the battery type and modeling, for its production and EoL, must also be key to the result. Some FFES companies declared recycling them through dedicated channels (Lelievre, 2019). These channels are still rare, do not provide information on the treatment and the fate of the battery components, and, in fine, on the environmental impact of this recycling activity. Battery LCI and their assessment globally suffer huge variabilities and uncertainties (Cox et al., 2018).

### 5.2.4 Static model limitations

This static study aims at assessing the impact of FFES penetration on GHG emissions in Paris. We thus compared emissions with and without FFES, over one year. But the FFES are not the only change in the Parisian transportation system. For instance, RATP, the Parisian public transportation operator, owns around 4700 buses and is conducting massive electrification of its rolling stock (RATP, 2019). The average impact of using a bus is thus changing over time. Other changes could affect other modes, and a dynamic model including these changes would be more accurate. Also, allocation factors for the infrastructure must be dynamic, i.e. calculated on a fixed time interval, for instance annually. Nevertheless, in the case of small modal changes like in this study, static allocation factors are accurate enough.

Finally, the limited availability of FFES has been pointed out as one of the limiting factors to its use (6t, 2019; Portland Bureau of Transportation, 2018). But FFES are all the more environmentally friendly as they reach long lifetime mileage. And multiplying the number of FFES could have the opposite effect. Resistant FFES designs to endure deterioration from weather conditions and rough shared usage, as well as vandalism, will be key to green performance, as well as the adequacy of the supply with the targeted demand through regulation.

### 5.3   Recommendations for micromobility policies

#### 5.3.1   Micromobility has high environmental potential but needs to be regulated

This paper shows that the FFES very likely caused an increase in the emissions from the mobility sector. Sensitivity and scenario analyses demonstrate that raising the FFES lifetime mileage would be positive though insufficient to reverse this conclusion, contrary to limiting servicing impacts severely. Moreover, cities with high electricity carbon intensities would more likely benefit from FFES from a climate change perspective than those with low carbon intensities such as in France, providing that the urban mobility largely relies on electrified public transportation with higher electricity consumption than the FFES. FFES and more generally microvehicles have the potential to reduce urban mobility carbon footprint. But they must be deployed carefully, with adequate regulations according to territorial characteristics, considering existing mobility systems and the energy sources they use. Without servicing impacts, the ES carbon footprint falls at 50 $gCO_2eq/km$. With longer though achievable lifetime mileages, resp. 5 200 to 15 600 km, it could drop resp. to 30 to 12 $gCO_2eq/km$. This impact is still higher than the metro, the RER, and walking in Paris, emitting respectively 7.6 g, 8.9 g, and 1.9 $gCO_2eq$/pkt. However, the ES has the potential to reduce the GHG emissions of the Parisian mobility (1) by solving the first-last mile issue that reduces mass transit adoption and (2) due to trip substitutions from more impacting modes. Outside the specific case of Paris or similar cities - in terms of population density, public transportation, and electricity carbon intensity -, FFES are likely to have a positive environmental effect.

### 5.3.2 *Shared mobility does not systematically mean reduced environmental impacts*

The common thinking is to assimilate the fact to share goods - for instance, vehicles - with a lower environmental footprint. Our case study shows how the results can be counterintuitive. If we consider taxi and ride-hailing as shared cars, what is debatable, the result is striking: it is in Paris the most emitting mode, with 299 $gCO_2eq$/pkt, when a private car emits 209 $gCO_2eq$/pkt, i.e. one third less. Shared bicycles emit more than private bicycles in Paris – 59 $gCO_2eq$/pkt against 15 g – partly due to the 30% electrification rate, rising their manufacturing stage impacts. The lifespan of these vehicles, and in general of shared vehicles, is also globally shorter, thus augmenting the environmental burden per kilometer traveled. This can be due to careless usages nay vandalism. Finally, servicing impacts - necessary to rebalance or charge the shared microvehicles - can be very impacting. It counts for around 50% of the FFES GHG emissions. Nevertheless, shared systems can be environmentally beneficial for occasional uses where a vehicle would have a low lifetime mileage. Moreover, these systems can trigger a ripple effect to reduce the mobility footprint by changing behaviors. First evidence shows that FFES have increased personal acquisitions of ES and other microvehicles in Paris. Considering their very low impact when their lifetime mileage is preserved, and providing that they do not substantially replace lower impact modes or induce extra trips, this phenomenon must generate environmental savings.

# 6 Conclusions

In this paper, we mathematically formalize a CLCA method to assess the environmental consequences of a change on a territory's mobility. We detail how to use this generic method on a specific case study relying on a mobility survey. In the case study, we calculate the impact of the FFES breakthrough on the GHG emissions from the Parisian mobility. The environmental performance of the twelve principal modes of transportation in Paris is calculated, and the contributions of the infrastructure and the vehicles on their complete life cycle are detailed. These intermediary results are now available for future transportation policy assessments using CLCA, in the context of Paris. The final results show that, in the case of Paris, the FFES disruption very likely caused an increase in GHG emissions from the mobility sector. Scenario analyses highlight that combining extended FFES lifetime mileages and limited servicing impacts is necessary to make FFES climate positive. Nevertheless, cities with high carbon intensity electricity mixes and a large mobility share supported by electric mass-rapid transit would more likely benefit from FFES in a climate change reduction perspective than those with low carbon intensity such as in France. More generally, FFES may have the potential to reduce urban mobility carbon footprint - but must be deployed carefully, with adequate regulations according to each city's characteristics - and micromobility certainly has a role to play towards a sustainable mobility.

**Acknowledgments**: We would like to thank Penelope Hawkins from Eurovia for her kind and patient proof-reading, as well as the four anonymous reviewers for their careful reading and astute suggestions.

**Funding source and role:** this study has been conducted under the research project ORNISIM funded by la fondation MAIF.

# SUPPLEMENTARY MATERIAL - CONSEQUENTIAL LCA FOR TERRITORIAL AND MULTIMODAL TRANSPORTATION POLICIES: METHOD AND APPLICATION TO THE FREE-FLOATING E-SCOOTER DISRUPTION IN PARIS

A. de Bortoli

## Table of contents

### 6.1   List of tables

### 6.2    List of figures

## 7    Survey methodology and site

We designed and undertook an extended survey among electric scooter (ES) users in Paris, conducted by face-to-face street interviews in May and June 2019. The questions focused on user profile (age, gender, occupation, etc.) and travel habits (frequency of use, distance covered, travel purpose, modal shift, etc.). The sample includes 459 valid answers and covers the 13 free-floating ES (FFES) operators: 53% of our respondents used Lime FFES for their FFES last trip, followed by Bird FFES with 22% of the FFES last trip, then by Bolt with 6% of the last trips. The other 10 operators account for 5% or less of the last trips.

To face the spatial heterogeneity challenge, we realized our survey in 7 preselected locations shown in Figure 7, ranging from pure residential areas in the inner suburbs (Issy-les-Moulineaux) to Paris CBD (La Defense). We also included two leisure sites (the banks of the Seine and the quays of St Martin canal), two major central public transportation hubs (Châtelet and République), and the area of Station F, the biggest startup campus in Paris.

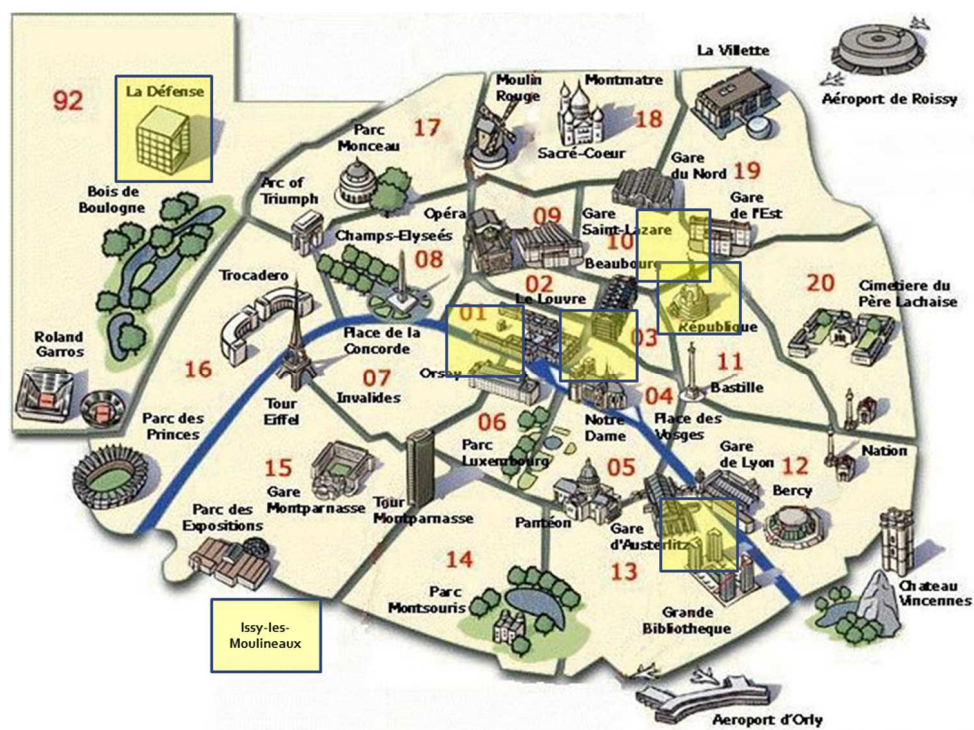

**Figure 7 Survey administering spots in yellow**

We also varied the time and day of the interviews to capture different user profiles. The sampling matrix is presented in Table 1 and follows the FFES usage distribution found by the company 6-t in an internet French survey (6t 2019b), among weekdays Vs weekend days as well as along the time of the day.

**Table 3 Schedule and area sampling matrix**

| Weekdays (Monday to Friday) | | | | | | |
|---|---|---|---|---|---|---|
| | 7-11h | 11-14h | 14-17h | 17-21h | 21-00h | Total |
| Issy-les-Moulineaux | 13 | 11 | 15 | 18 | 4 | 63 |
| La Défense | 13 | 11 | 15 | 18 | 4 | 63 |
| Banks of the Seine | 13 | 11 | 15 | 18 | 4 | 63 |
| Quays of Canal St Martin | 13 | 11 | 15 | 18 | 4 | 63 |
| Châtelet | 13 | 11 | 15 | 18 | 4 | 63 |
| Station F area | 13 | 11 | 15 | 18 | 4 | 63 |
| République | 13 | 11 | 15 | 18 | 4 | 63 |
| Total | 75 | 68 | 90 | 105 | 26 | 366 |
| Weekend days (Saturday and Sunday) | | | | | | |
| | 7-11h | 11-14h | 14-17h | 17-21h | 21-00h | Total |
| Banks of the Seine | 3 | 9 | 21 | 10 | 3 | 49.14 |
| Quays of Canal St Martin | 3 | 9 | 21 | 10 | 3 | 49.14 |
| Châtelet | 3 | 9 | 21 | 10 | 3 | 49.14 |
| Station F area | 3 | 9 | 21 | 10 | 3 | 49.14 |
| République | 3 | 9 | 21 | 10 | 3 | 49.14 |
| Total | 17 | 44 | 103 | 52 | 17 | 234 |

*The sample statistics are represented in*

*Table 4. Respondents were 2/3 of men and 1/3 of women, while the average Parisian population is gender-balanced (INSEE 2019). 97% of the respondents live in France: 56% in Paris inner-city, 13% in the close suburbs, 11% in the rest of the Parisian region, and 20% elsewhere in France. The visiting users (3%) mostly came from Italy (43%), the USA (21%), and Switzerland (14%). We note here that the percentage of foreign tourists is surprisingly low even though the questionnaire was administrated in both French and English. We believe that the non-response rate mainly concerned this market segment. Compared to the average Parisian population (INSEE 2019), some age classes are over-represented – namely people under 30 (46% instead of 37%) - while other are largely under-represented – people over 45 (8% against 32%) and people under 15 years of age (less than 2% against 14%).*

**Table 4 Survey sample**

| Variable | Type | Summary statistics (%) | Description |
|---|---|---|---|
| Gender | | F(1)= 68 | 1 = "male" |
| | | F(2)= 32 | 2= "female" |
| Place of living | | F(1)=97 | 1=France |
| | | F(2)=3 | 2=Others |
| Age | | F(1)= 8 | 1="12 to 17" |
| | | F(2)= 38 | 2="18 to 24" |
| | Discrete | F(3)=25 | 3="25 to 29" |
| | | F(4)=14 | 4="30 to 34" |
| | | F(5)=7 | 5="35 to 44" |
| | | F(6)=8 | 6=">=45" |
| Monthly income | | F(1)= 38 | 1="none" |
| | | F(2)= 17 | 2="Less than 900€" |
| | | F(3)= 13 | 3="From 901 to 2000 €" |
| | | F(4)= 18 | 4=" From 2001 to 3000 €" |
| | | F(5)=16 | 5=" From 3001 to 5000 €" |
| | | F(6)=5 | 6=" More than 5000 €" |
| Occupation category | | F(1)= 30 | =1 if "Head of a company" or "Executive, intellectual employment" ; =2 if "intermediate employment" or "Craftsperson, retail trader"; =3 if "worker" or "Employee" ; 4= "none" (student, at school, retired) |
| | | F(2)=11 | |
| | | F(3)= 18 | |
| | | F(4)= 42 | |

| Last degree | | F(1)= 11 | 1 if none or before high school diploma |
|---|---|---|---|
| | | F(2)= 11 | 2 if professional degree under high school |
| | | F(3)= 25 | 3="general high school diploma" |
| | | F(4)= 21 | 4 = "some college credit" |
| | | F(5)=30 | 5 if "master's degree" and above |
| | | F(6)=3 | 6 if others or did not answer |
| Household composition | | F(1)= 45 | 1= "living alone" |
| | | F(2)= 22 | 2= "childless couple" |
| | | F(3)= 17 | 3="Couple with child(ren)" |
| | | F(4)=3 | 4="Alone with one or more children" |
| | | F(5)=12 | 5="with roommate(s)" |
| ES equipment | Binary | F(1)=7% | 1=ES owner; 2=exclusively FFES user |

# 8   FFES last trip characteristics, modal speeds and walking access assumptions

**Table 5 Modal shifts characteristics: trip-based modal shift percentages based on our survey on, walking access assumptions and modal average speed**

| Mode | Unweighted trip-based modal shift (%) | Accessing walking distance (m) | Data source | Average/ commercial speed (km/h) | Previous data source |
|------|------|------|------|------|------|
| Walk | 34.6 | 0 | | 4.7 | Calculated by authors from 20 itineraries in Paris, using Google Maps |
| Personal bicycle | 3.9 | 0 | | 15.0 | Authors' estimation |
| Shared bicycle | 3.5 | 400 | Assumption | 15.0 | Authors' estimation |
| Shared 2-wheeler | 1.5 | 300 | Calculated from Parisian data (6t 2019a) | 24.0 | Considered equal to a motorbike (Andy 2018) |
| Personal 2-wheeler | 3.7 | 0 | | 24.0 | Calculated by authors from a Parisian experience (Andy 2018) |
| E-scooter | 0.2 | 0 | | 17.0 | Authors (survey statistics) |
| Car | 3.7 | 0 | | 15.0 | (Ville de Paris 2016) |
| Taxi & ride-hailing | 6.1 | 0 | | 16.8 | Average speed based on GPS data from a leading taxi company (Dell'oro 2014) |
| Bus | 11.5 | 400 | Assumption | 12.5 | (Carsuzaa 2013); (ortferroviaire 2018) |
| Metro | 23.3 | 1200 | Assumption | 30.0 | (Carsuzaa 2013) |
| RER | 2.2 | 1200 | Assumption | 49.5 | (RATP 2017) |
| Streetcar | 0.4 | 1200 | Assumption | 19.0 | (ortferroviaire 2018);(Carsuzaa 2013) |

| Induced trips | 5.4 | NA | | NA | NA |
|---|---|---|---|---|---|

# 9   Life Cycle Inventories

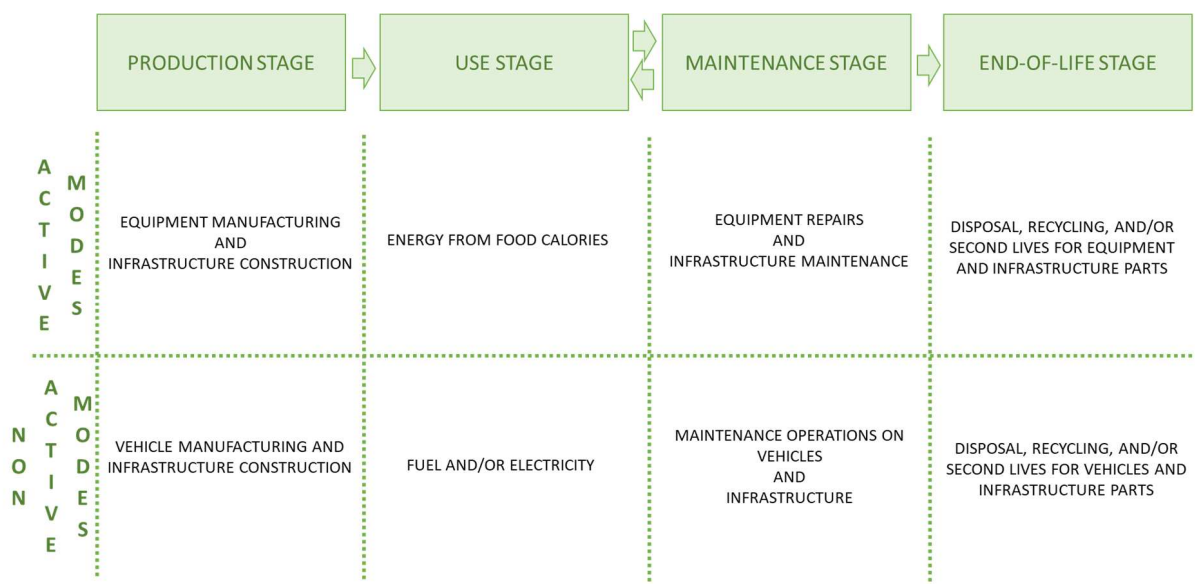

**Figure 8 Detail of the respective life cycles of active Vs non-active modes**

The free-floating electric scooter Life Cycle Inventories are detailed in Table 3. The material inventory comes from Hollingsworth et al. (2019b) with the following changes: the used lithium-ion battery weight is scaled down to the weigh of the non-used battery (1.159 kg, i.e. - 100 g than in the seminal study), the heat (other than natural gas) is considered consumed "at central or small scale" (no ecoinvent process at the district or industrial scale), the recycling LCI of the FFES is scaled down from the recycling process for one bike (of 22 kg) based on the FFES weigh (12.5 kg), and the battery LCI is the one from ecoinvent instead of coming from Ellingsen et al. (2014).

The transportation stage inventories consider the following itinerary: the FFES come from Shenzhen by train (average Chinese market trains from Shenzhen to Alanshankou in China (around 4 500 km); then diesel trains in Kazakhstan (Alanshankou, China to Troïskt, Russia: around 2 400 km) and the rest of the trip in electric trains: 1 500 km from Troïskt, Russia, to Lodz, Poland) to the city of Lodz in Poland, before being transported by Euro 5 trucks to Paris (1 500 km).

Based on our survey, ES have a battery autonomy of 20 km in Paris conditions instead of the theoretical 30 km given by the constructor. It requires 335 Wh per full charge.

**Table 6 LCI for one FFES, except for the use stage**

| Flow | Amount | Unit |
|---|---|---|
| aluminium alloy, AlMg3 \| market for aluminium alloy, AlMg3, GLO | 5,731 | kg |
| aluminium, cast alloy \| market for aluminium, cast alloy, GLO | 0,256 | kg |
| battery cell, Li-ion \| battery cell production, Li-ion, CN | 1,159 | kg |
| charger, for electric scooter \| charger production, for electric scooter, GLO | 0,385 | kg |
| electric motor, for electric scooter \| electric motor production, for electric scooter, GLO | 1,187 | kg |
| electricity, medium voltage, aluminium industry \| market for electricity, medium voltage, aluminium industry, CN | 6,89 | kWh |
| heat, central or small-scale, other than natural gas \| market for heat, central or small-scale, other than natural gas, ROW | 0,193 | MJ |
| heat, district or industrial, natural gas \| market for heat, district or industrial, natural gas, ROW | 13,6 | MJ |
| light emitting diode \| light emitting diode production, GLO | 0,016 | kg |
| municipal solid waste \| market for municipal solid waste, ROW | -4,5 | kg |
| polycarbonate \| market for polycarbonate, GLO | 0,274 | kg |
| powder coat, aluminium sheet \| market for powder coat, aluminium sheet, GLO | 0,35 | m2 |
| printed wiring board, surface mounted, unspecified, Pb containing \| printed wiring board production, surface mounted, unspecified, Pb containing, GLO | 0,059 | kg |
| steel, low-alloyed \| market for steel, low-alloyed, GLO | 1,349 | kg |
| synthetic rubber \| market for synthetic rubber, GLO | 1,185 | kg |
| tap water \| market for tap water, ROW | 0,744 | kg |
| transistor, wired, small size, through-hole mounting \| market for transistor, wired, small size, through-hole mounting, GLO | 0,062 | kg |
| transport, freight train \| market for transport, freight train, CN | 56,25 | t*km |
| transport, freight train \| transport, freight train, diesel, CN | 30 | t*km |
| transport, freight train \| transport, freight train, electricity, ROW | 18,75 | t*km |
| transport, freight, lorry, unspecified \| transport, freight, lorry, all sizes, EURO5 to generic market for transport, freight, lorry, unspecified, ROW | 18,75 | t*km |
| used electric bicycle \| market for used electric bicycle, GLO | -0,568 | Item(s) |
| used Li-ion battery \| market for used Li-ion battery, GLO | -1,159 | kg |
| wastewater, average \| market for wastewater, average, GLO | -0,0007 | m3 |
| welding, arc, aluminium \| market for welding, arc, aluminium, GLO | 0,75 | m |

**Table 7 Life cycle model for the other vehicles and equipment: details on ecoinvent processes and quantities for each stage of the life cycle**

| Stage | Production | | Maintenance | | EoL | | Lifetime mileage (km) |
|---|---|---|---|---|---|---|---|
| Mobility macro-process | Process | Quantity | Process | Quantity | Process | Quantity | |
| Walk | Shoes neglected | | Neglected | | | | |
| Personal bicycle | "market for bicycle, GLO" | 20.6/17 item(s) | "market for maintenance, bicycle, GLO" | 20.6/17 item(s) | Included in production | | 15,000 |
| Shared bicycle | "market for bicycle GLO" | 20.6/17 item(s) | "market for maintenance, bicycle, GLO" | 20.6/17 item(s) | Included in production | | 4,500 |
| | "electric motor, vehicle" | 1/3*2.2 kg | No maintenance | | No process | | |
| | "battery, Li-ion, rechargeable, prismatic, GLO" | 1/3*3.9 kg | No maintenance | | "market for used Li-ion battery, GLO" | -1/3*3.9 kg | |

| | | | | | | |
|---|---|---|---|---|---|---|
| **Shared 2-wheelers** | "market for electric scooter, without battery, GLO" | 112 kg | "market for maintenance, electric scooter, without battery, GLO | 1 item(s) | Included in production | | 50,000 |
| | "market for battery cell, Li-ion, GLO" | 8 kg | No maintenance | | "market for used Li-ion battery, GLO" | -8 kg | |
| **Personal 2-wheeler** | market for motor scooter, 50 cubic cm engine, GLO | 1.5 item(s) | "market for maintenance, motor scooter, GLO | 1 item(s) | Included in production | | 50,000 |
| **Personal car** | "market for passenger car, diesel, GLO" | 0.61*1278 | "maintenance, passenger car, RER" | 1 item(s) | dismantling included in production | | 200,000 |
| | "market for passenger car, petrol/natural gas, GLO" | 0.39*1278 | | | | | |
| **Taxi & ride-hailing** | "market for passenger car, diesel, GLO" | 0.82*1400 kg | "maintenance, passenger car, RER" | 0.82+0.07 item(s) | dismantling included in production | | |
| | "market for passenger car, petrol/natural gas, GLO" | 0.07*1400 kg | | | | | |
| | "market for electric car without battery, GLO" | 0.11*1400 kg | "maintenance, passenger car, electric, without battery, GLO" | 0.11 item(s) | dismantling included in production | | |
| | "battery cell, Li-ion, GLO | 0.11*250 kg | No maintenance for battery | | "market for used Li-ion battery, GLO" | -0.11*250 kg | |
| **Bus** | ecoinvent - "Bus production, RER" | 1 item(s) | "Maintenance, bus, RoW" | 1 item(s) | Included in production | | 480,000 |

| Metro | "train production, passenger, regional, RoW | 131/171 item(s) | "Maintenance, locomotive, RER" | 131/171 item(s) | Included in production | 6,000,000 |
| RER | "train production, passenger, regional, RoW" | 230/171 item(s) | "Maintenance, locomotive, RER" | 230/171 item(s) | Included in production | 6,000,000 |
| Streetcar | "tram production, RER) | 52/21 item(s) | "maintenance, tram, RoW" | 52/21 item(s) | Included in production | 1,120,000 |

**Table 8 Emissions and consumptions of the different vehicles, and sources of the data**

| | Occupancy (pax/veh) | Exhaust emissions (gCO2 per pkt) | Consumption (kWh/pkt or kg/pkt) | Source | Comments |
|---|---|---|---|---|---|
| RER | 441 | 5.40E+00 | 1.13E-01 | | RATP in SNCF 2019 |
| Metro | 161 | 3.40E+00 | 7.08E-02 | | RATP in SNCF 2019 |
| Streetcar | 113 | 2.80E+00 | 5.83E-02 | | RATP in SNCF 2019 |
| Diesel bus | 17 | 9.35E+01 | 1.17E-02 | | RATP in SNCF 2019 |
| Gasoline private car | 1.3 | 1.36E+02 | 4.34E-02 | ADEME 2019 | Direct emissions |
| Diesel private car | 1.3 | 1.28E+02 | 4.19E-02 | ADEME 2019 | Direct emissions |
| Electric Sedan taxi | 0.85 | NA | 2.94E-01 | Estimated | Electricity emission |
| Gasoline taxi | 0.85 | 2.21E+02 | 7.05E-02 | ADEME 2019 | Direct emissions, 6-10 power horses |
| Diesel taxi | 0.85 | 2.06E+02 | 6.71E-02 | ADEME 2019 | Direct emissions, 6-10 power horses |
| Electric bicycle | 1 | NA | 1.00E-02 | | Leuenberger and Frischknecht 2010 |
| Electric motor scooter | 1 | NA | 3.00E-02 | | Leuenberger and Frischknecht 2010 |

| Gasoline motor scooter | 1 | 6.95E+01 | 3.31E-02 | ADEME 2019 - (47%/2 >250cm3-short distance, 47%/2 <=250cm3 and >125cm3, 38% <=125cm3 and >50cm3, 15%=50cm3 (DRIEA - STIF 2013)) |

**Table 9 Life cycle inventories for the three street infrastructure types**

|  | Cycle lane | Sidewalk | Pavement |
|---|---|---|---|
| **Functional Unit** | 1 linear meter over one year | 1 m² over one year | 1 m2 over one year |
| **Binder (kg)** | 2,47E-01 | 2,88E-01 | 7,05E-01 |
| **Gravel (kg)** | 9,37E+00 | 6,44E+00 | 1,77E+01 |
| **Concrete block (kg)** | 5,40E+00 | 0,00E+00 | 0,00E+00 |
| **Truck transportation (tkm)** | 7,51E-01 | 3,36E-01 | 9,21E-01 |
| **HMA manufacturing (kg)** | 4,11E+00 | 2,88E+00 | 1,11E+01 |

*Nb: The density considered are 1.54 for the gravel, and 2.3 for the compacted HMA*

**Table 10 Life cycle inventory for the manufacture of one metric ton of hot mix asphalt in France**

| Input flow | Unit | Amount |
|---|---|---|
| electricity, medium voltage \| market for electricity, medium voltage | kWh | 7.25 |
| heat, central or small-scale, other than natural gas \| heat production, light fuel oil, at boiler 100kW, non-modulating | kWh | 2.08 |
| heat, central or small-scale, other than natural gas \| heat production, lignite briquette, at stove 5-15kW | kWh | 0.7 |
| heat, district or industrial, natural gas \| heat production, natural gas, at industrial furnace low-NOx >100kW | kWh | 47.6 |
| heat, district or industrial, other than natural gas \| heat production, light fuel oil, at industrial furnace 1MW | kWh | 1.4 |
| heavy fuel oil, burned in refinery furnace \| heavy fuel oil, burned in refinery furnace | kWh | 21 |
| tap water \| market for tap water | kg | 700 |
| wastewater, average \| treatment of wastewater, average, capacity 1E9l/year | m3 | 2.32E-03 |
| Water, well, in ground | m3 | 9.53E-05 |
| **Output flow** | **Unit** | **Amount** |
| BOD5, Biological Oxygen Demand | kg | 4.84E-05 |
| COD, Chemical Oxygen Demand | kg | 8.26E-04 |
| Hot Mix Asphalt manufacture FR | t | 1 |
| Hydrocarbons, unspecified | kg | 5.82E-06 |
| Nitrate | kg | 3.57E-06 |
| Suspended solids, unspecified | kg | 7.49E-04 |

# 10 Transportation system data for the city of Paris

**Table 11 Volume of each kind of infrastructural network and related mobility in Paris**

| | Annual mobility (pkt) | | Annual offer (vkt) | | Network |
|---|---|---|---|---|---|
| **Streetcar** | 3,27E+08 | OMNIL 2019 corrected by the ratio of passengers on the Parisian lines | 2,90E+06 | OMNIL 2019, only lines T3a and b | 29 km |
| **Metro** | 8,11E+09 | OMNIL 2019 | 5,05E+07 | OMNIL 2019 (including all the traffic on the RATP network) | 206 km |
| **RER** | 5,51E+09 | OMNIL 2019 (RER trains operated by the RATP only) | 1,25E+07 | OMNIL 2019 | 80 km |
| **Bus** | 8,27E+08 | OMNIL 2019 (bus in Paris + night's service) | 4,78E+07 | OMNIL 2019, including night busses | 75 ha of dedicated lanes |
| **Taxis** | 2,33E+09 | EGT 2010 | 1,86E+09 | EGT 2010 | 1220 ha of pavement + 198 ha of street parking lots |
| **Personal cars** | | | | | |
| **Motorized two-wheelers** | 3,93E+08 | EGT 2010 | 3,93E+08 | EGT 2010 | |
| **Bicycles** | 3,72E+08 | EGT 2010 + Heran 2018 | 3,72E+08 | EGT 2011 | 85 + 21 ha of lanes + 7 ha of parking lots |
| **Walking** | 7,49E+08 | EGT 2010 (underestimation) | 7,49E+08 | EGT 2012 | 1200 ha |

**Table 12 Traffic (vkt and pkt) in 2017 and 2018 in Paris inner-city**

| Mode | 2017 PKT | 2017 VKT | 2018 PKT | 2018 VKT |
|---|---|---|---|---|
| **RER** | 5,51E+09 | 1,25E+07 | 5,48E+09 | 1,25E+07 |
| **Metro** | 8,11E+09 | 5,05E+07 | 7,92E+09 | 4,99E+07 |
| **Streetcar** | 3,27E+08 | 2,90E+06 | 3,26E+08 | 2,90E+06 |
| **Bus** | 8,27E+08 | 4,78E+07 | 8,03E+08 | 4,71E+07 |
| **Private car** | 2,33E+09 | 1,86E+09 | 2,31E+09 | 1,85E+09 |
| **Taxi and ride-hailing** | | | | |
| **Shared motor scooter** | 3,93E+08 | 3,93E+08 | 3,74E+08 | 3,74E+08 |
| **Private motor scooter** | | | | |
| **Shared e-scooter** | 0,00E+00 | 0,00E+00 | 2,41E+08 | 2,41E+08 |
| **Shared bicycle** | 3,72E+08 | 3,72E+08 | 3,37E+08 | 3,37E+08 |
| **Private bicycle** | | | | |
| **Walking** | 7,49E+08 | 7,49E+08 | 6,87E+08 | 6,87E+08 |

| Other annual traffics | Annual vkt (2014) | Source |
|---|---|---|
| **Trucks** | 7,40E+05 | Airparif 2018 |
| **Commercial vehicles** | 4,80E+06 | Airparif 2018 |

# 11 Paris infrastructure allocation factors

**Table 13 Final allocation factors in 2017 and 2018**

| | 2017 allocation factor | 2018 allocation factor |
|---|---|---|
| **RER** | 1,81E-10 | 1,83E-10 |
| **Metro** | 1,23E-10 | 1,26E-10 |
| **Streetcar** | 3,06E-09 | 3,07E-09 |
| **Bus on bus lane** | 1,21E-09 | 1,24E-09 |
| **Bus on other roads** | 4,34E-10 | 4,39E-10 |
| **Taxi and ride-hailing** | 4,34E-10 | 4,39E-10 |
| **Private car** | 4,33E-10 | 4,39E-10 |
| **Truck** | 4,34E-10 | 4,39E-10 |
| **Motorized two-wheeler** | 4,33E-10 | 4,39E-10 |
| **Bicycle and FFES** | 2,68E-09 | 1,73E-09 |
| **Pedestrian** | 1,33E-09 | 1,45E-09 |

# 12 Results - carbon footprints

**Table 14 Life cycle carbon footprint for the main modes of transportation in Paris, in kgCO$_2$eq/pkt**

| Mode | Vehicle | Use stage | Servicing | Infrastructure | Total |
|------|---------|-----------|-----------|----------------|-------|
| **Walking** | 0.00E+00 | 0.00E+00 | 0.00E+00 | 2.23E-03 | 2.23E-03 |
| **Metro** | 6.70E-04 | 4.45E-03 | 0.00E+00 | 2.43E-03 | 7.55E-03 |
| **RER** | 4.29E-04 | 7.00E-03 | 0.00E+00 | 1.37E-03 | 8.79E-03 |
| **Private bicycle** | 1.41E-02 | 0.00E+00 | 0.00E+00 | 1.13E-03 | 1.52E-02 |
| **Streetcar** | 5.65E-03 | 3.71E-03 | 0.00E+00 | 1.08E-02 | 2.02E-02 |
| **Shared motor scooter** | 1.98E-02 | 1.91E-03 | 0.00E+00 | 6.03E-03 | 2.77E-02 |
| **Shared bicycle** | 5.75E-02 | 2.15E-04 | 2.55E-04 | 1.13E-03 | 5.91E-02 |
| **Shared e-scooter** | 5.56E-02 | 1.21E-03 | 5.14E-02 | 1.13E-03 | 1.09E-01 |
| **Bus** | 1.55E-02 | 1.11E-01 | 0.00E+00 | 7.08E-03 | 1.33E-01 |
| **Private motor scooter** | 2.21E-02 | 1.07E-01 | 0.00E+00 | 6.03E-03 | 1.35E-01 |
| **Private car** | 2.91E-02 | 1.74E-01 | 0.00E+00 | 6.03E-03 | 2.09E-01 |
| **Taxi and ride-hailing** | 4.97E-02 | 2.44E-01 | 0.00E+00 | 6.03E-03 | 2.99E-01 |

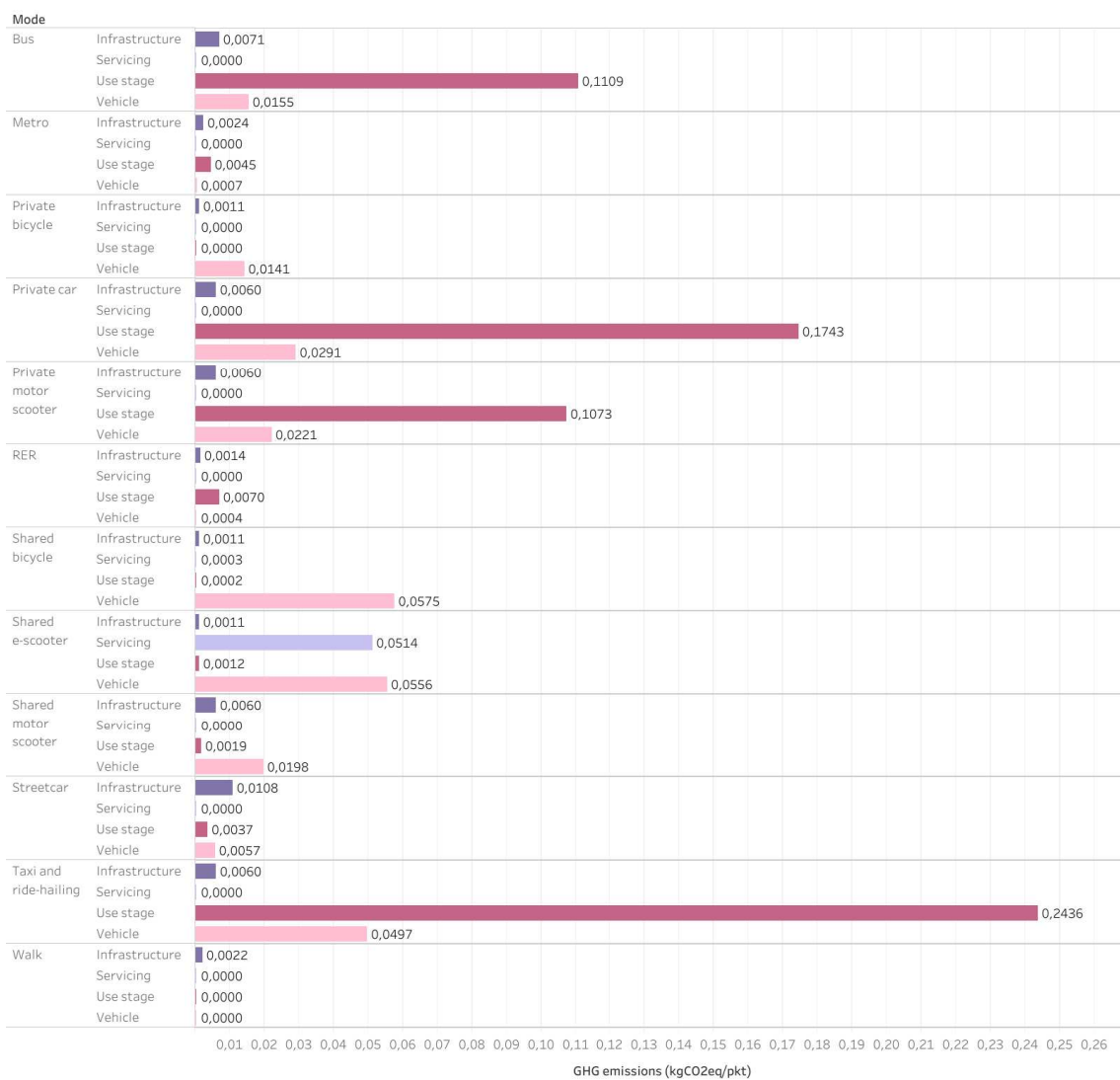

**Figure 9 Life cycle carbon footprint for the main modes of transportation in Paris, in kgCO2eq/pkt**

**Table 15 Life cycle carbon footprint for different servicing scenarios**

| Scenario name | Charger type | Transpor-tation distance | Mode | Number of ES | km/ES.day | Unitary emission impact (kgCO2e/vkt) | Servicing total impact (kgCO2e/pkt) | FFES marginal impact (1000 tCO2e) |
|---|---|---|---|---|---|---|---|---|
| **LCV 90 km 50 ES** | Pro | 90 | LCV | 50 | 1,8 | 6,30E-01 | 1,03E-01 | 2.54 E+01 |
| **LCV 90 km 100 ES** | Pro | 90 | LCV | 100 | 0,9 | 6,30E-01 | 5,15E-02 | 1.29 E+01 |

| Juicer 10 km car | Entrepreneur | 10 | Car | 11 | 0,909091 | | 1,69E-02 | |
| | | | | | | 2,04E-01 | | 4.57E+00 |
| Swappable battery 90 km car | Pro | 90 | Car | 200 | 0,45 | | 8,35E-03 | |
| | | | | | | 2,04E-01 | | 2.51E+00 |
| Swappable battery 45 km car | Pro | 45 | Car | 200 | 0,225 | | 4,17E-03 | |
| | | | | | | 2,04E-01 | | 1.51E+00 |
| Riding juicer | Side -activity | 4 | ES | 2 | 0,5 | | 1,17E-03 | |
| | | | | | | 2,57E-02 | | 7.80E-01 |
| Walking juicer | Side- activity | 3 | Walk | 2 | 1,5 | | 2,42E-04 | |
| | | | | | | 1,93E-03 | | 5.57E-01 |

We consider that each FFES travels 11 km per day (Hollingsworth, Copeland, and Johnson 2019a)

**Table 16 Carbon intensity of electricity mixes (high and low voltages) in different countries using CML 2001 characterization factors and ecoinvent V3.2, consequential system**

| Country | ecoinvent code | Market, high voltage - GHG emissions (kgCO2e/kWh) |
| --- | --- | --- |
| Switzerland | CH | 8.87E-03 |
| Norway | NO | 1.45E-02 |
| France | FR | 6.36E-02 |
| Belgium | BE | 1.46E-01 |
| Northeast Power Coordinating Council, US part only | NPCC, US only | 1.82E-01 |
| Spain | ES | 4.16E-01 |
| Average Europe | RER | 4.31E-01 |
| Italy | IT | 5.16E-01 |
| Average US | US | 6.70E-01 |
| Florida | FRCC | 6.73E-01 |

| Germany | DE | 9.17E-01 |
| China | CN | 1.16E+00 |